\documentclass[10pt,conference]{IEEEtran}
\usepackage{cite}
\usepackage{amsmath,amssymb,amsfonts}
\usepackage{graphicx}
\usepackage{textcomp}
\usepackage[hyphens]{url}
\usepackage{fancyhdr}
\usepackage{xcolor}
\usepackage{listings}

\usepackage{mathptmx} 
\usepackage[normalem]{ulem}
\usepackage{wrapfig}
\usepackage{mwe}
\usepackage{pifont}
\usepackage[T1]{fontenc}  
\usepackage{balance}
\usepackage[noend]{algorithmic}
\usepackage{algorithm}
\usepackage{booktabs, multirow} 
\usepackage{soul}
\usepackage{changepage,threeparttable} 
\usepackage[colorinlistoftodos,prependcaption]{todonotes}
\usepackage{xspace}
\usepackage{tabulary}
\usepackage[bookmarks=true,breaklinks=true,letterpaper=true,colorlinks,linkcolor=black,citecolor=blue,urlcolor=black]{hyperref}
\usepackage[]{rotating}

\newcommand{\gpic}[1]{MVE}
\newcommand{\Gpic}[1]{MVE}
\newcommand{\ignore}[1]{}
\definecolor{mygreen}{RGB}{0, 65, 0}
\definecolor{myblue}{RGB}{14, 0, 210}
\newcommand{\blue}[1]{\textcolor{myblue}{#1}}

\newcommand{\cmark}{\ding{51}}%
\newcommand{\xmark}{\ding{55}}%

\newcommand{\vrotate}[1]{\rotatebox[origin=c]{90}{#1}}

\definecolor{mGreen}{rgb}{0,0.6,0}
\definecolor{mGray}{rgb}{0.5,0.5,0.5}
\definecolor{mPurple}{rgb}{0.58,0,0.82}
\definecolor{backgroundColour}{rgb}{1,1,1}
\definecolor{mygreen}{RGB}{0, 100, 30}
\definecolor{myred}{RGB}{200, 00, 00}

\lstdefinestyle{CStyle}{
    backgroundcolor=\color{backgroundColour},   
    commentstyle=\color{mGreen},
    keywordstyle=\color{magenta},
    numberstyle=\tiny\color{mGray},
    stringstyle=\color{mPurple},
    basicstyle=\footnotesize,
    breakatwhitespace=false,         
    breaklines=true,                 
    captionpos=b,                    
    keepspaces=true,                 
    numbers=left,                    
    numbersep=5pt,                  
    showspaces=false,                
    showstringspaces=false,
    showtabs=false,                  
    tabsize=2,
    frame=single,
    linewidth=.48\textwidth,
    language=C
}

\newcommand{\hpcayear}{2025}

\newcommand{\hpcasubmissionnumber}{361}
\title{Multi-Dimensional Vector ISA Extension \\ for Mobile In-Cache Computing}

\def\hpcacameraready{} 

\newcommand\hpcaauthors{Alireza Khadem$^\dagger$, Daichi Fujiki$^\ddagger$, Hilbert Chen$^\dagger$, Yufeng Gu$^\dagger$, Nishil Talati$^\dagger$, Scott Mahlke$^\dagger$, and Reetuparna Das$^\dagger$}
\newcommand\hpcaaffiliation{University of Michigan$^\dagger$, Institute of Science Tokyo$^\ddagger$}
\newcommand\hpcaemail{\{arkhadem,cyaa,yufenggu,talatin,mahlke,reetudas\}@umich.edu, dfujiki@artic.iir.isct.ac.jp}

\author{
  \ifdefined\hpcacameraready
    \IEEEauthorblockN{\hpcaauthors{}}
      \IEEEauthorblockA{
        \hpcaaffiliation{} \\
        \hpcaemail{}
      }
  \else
    \IEEEauthorblockN{\normalsize{HPCA \hpcayear{} Submission
      \textbf{\#\hpcasubmissionnumber{}}} \\
      \IEEEauthorblockA{
        Confidential Draft \\
        Do NOT Distribute!!
      }
    }
  \fi 
}

\fancypagestyle{camerareadyfirstpage}{%
  \fancyhead{}
  
  \fancyhead[C]{
    \ifdefined\aeopen
    \parbox[][12mm][t]{13.5cm}{\hpcayear{} IEEE International Symposium on High-Performance Computer Architecture (HPCA)}    
    \else
      \ifdefined\aereviewed
      \parbox[][12mm][t]{13.5cm}{\hpcayear{} IEEE International Symposium on High-Performance Computer Architecture (HPCA)}
      \else
      \ifdefined\aereproduced
      \parbox[][12mm][t]{13.5cm}{\hpcayear{} IEEE International Symposium on High-Performance Computer Architecture (HPCA)}
      \else
      \parbox[][0mm][t]{13.5cm}{\hpcayear{} IEEE International Symposium on High-Performance Computer Architecture (HPCA)}
    \fi 
    \fi 
    \fi 
    \ifdefined\aeopen 
      \includegraphics[width=12mm,height=12mm]{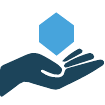}
    \fi 
    \ifdefined\aereviewed
      \includegraphics[width=12mm,height=12mm]{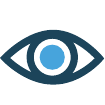}
    \fi 
    \ifdefined\aereproduced
      \includegraphics[width=12mm,height=12mm]{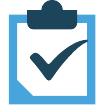}
    \fi
  }
  \fancyfoot[C]{}
}
\begin{document}

\maketitle

\ifdefined\hpcacameraready 
  \thispagestyle{camerareadyfirstpage}
  \pagestyle{empty}
\else
  \thispagestyle{plain}
  \pagestyle{plain}
\fi

\newcommand{\hpcaheight}{0mm}
\ifdefined\eaopen
\renewcommand{\hpcaheight}{12mm}
\fi

\renewcommand{\thefootnote}{}
\footnotetext{\copyright 2025 IEEE.  Personal use of this material is permitted.  Permission from IEEE must be obtained for all other uses, in any current or future media, including reprinting/republishing this material for advertising or promotional purposes, creating new collective works, for resale or redistribution to servers or lists, or reuse of any copyrighted component of this work in other works.}
\renewcommand{\thefootnote}{\arabic{footnote}} 



\begin{abstract}

In-cache computing technology transforms existing caches into long-vector compute units and offers low-cost alternatives to building expensive vector engines for mobile CPUs. Unfortunately, existing long-vector Instruction Set Architecture (ISA) extensions, such as RISC-V Vector Extension (RVV) and Arm Scalable Vector Extension (SVE), provide only one-dimensional strided and random memory accesses. While this is sufficient for typical vector engines, it fails to effectively utilize the large Single Instruction, Multiple Data (SIMD) widths of in-cache vector engines. This is because mobile data-parallel kernels expose limited parallelism across a single dimension.

Based on our analysis of mobile vector kernels, we introduce a long-vector \underline{M}ulti-dimensional \underline{V}ector ISA \underline{E}xtension (\gpic{}) for mobile in-cache computing. \gpic{} achieves high SIMD resource utilization and enables flexible programming by abstracting cache geometry and data layout. The proposed ISA features multi-dimensional strided and random memory accesses and efficient dimension-level masked execution to encode parallelism across multiple dimensions. Using a wide range of data-parallel mobile workloads, we demonstrate that \gpic{} offers significant performance and energy reduction benefits of 2.9$\times$ and 8.8$\times$, on average, compared to the SIMD units of a commercial mobile processor, at an area overhead of 3.6\%.

\end{abstract}
\section{Introduction}~\label{sec:intro}

\vspace{-4mm}

Data-parallel kernels are ubiquitous in various mobile application domains, from operating systems (\textit{e.g.}, Android) to web applications (\textit{e.g.}, Google Chrome)~\cite{swan}. These kernels typically consist of tight loops that interleave scalar and vector processing.
Single Instruction Multiple Data (SIMD) pipelines integrated with a CPU core are employed to exploit these applications' \textit{fine-grain data-level parallelism}.
SIMD execution amortizes the cost of instruction fetch and decode to improve the performance and energy consumption of data-parallel applications, \textit{i.e.}, two primary concerns of resource-limited mobile devices.

While mobile applications benefit from vector architectures, scaling up vector register files and ALUs incurs significant area and power overheads to the mobile processors. Hence, all Arm mobile processor implementations are restricted to 128-bit vector units. Alternatively, various In-SRAM computing schemes~\cite{computecaches, neuralcache, bitprudent, bsbppincache, bcam, eve} repurpose cache arrays to build high-throughput and power-efficient vector engines with minimal area overhead. For instance, half the size  of a small private L2 cache (256 KB) in a mobile CPU can be transformed into an 8192-lane bit-serial vector engine with a 3.5$\%$ area overhead~\cite{dualitycache} to the die. The cache SRAM arrays provide compute capability and also act as vector registers. Furthermore, they function as regular caches without performance overhead when not used for In-SRAM computing.

Existing long-vector ISA extensions (\textit{e.g.}, RISC-V RVV~\cite{riscv} and Arm SVE~\cite{armsve}) provide one-dimensional strided and random memory accesses. While sufficient for typical vector engines (128${-}$1024 bits), these fall short of effectively utilizing long-vector in-cache vector engines with large SIMD width (8192$\times$32 bits). Using conventional long-vector ISAs with only one-dimensional memory accesses requires several vector manipulation instructions to load/store and unpack/pack multiple dimensions of the data structures to a single long-vector register.
This substantially increases the instruction count and core-cache communication, under-utilizing the in-cache SIMD lanes.

This issue is compounded by the fact that mobile vector kernels expose \textit{limited one-dimensional (1D) parallelism}.
Our evaluation shows an average of 635 1D parallelism in a vector benchmark suite for mobile processors~\cite{swan}. This benchmark encompasses mobile applications from various domains, such as Chromium~\cite{chromium} (browser and OS), Android~\cite{android} (OS), PDFium~\cite{pdfium} (PDF rendering engine), and WebRTC~\cite{webrtc} (real-time voice/text/video communication APIs for conferencing platforms such as Zoom, Microsoft Teams, Slack, or Google Meet). For example, the audio processing module of the WebRTC (\textit{webaudio}~\cite{webaudio}) simultaneously processes multiple audio chunks, each containing different channels of 128 audio samples, which limits the 1D parallelism to only 128 elements.


This observation motivates \gpic{}, a long-vector \underline{M}ulti-dimensional \underline{V}ector ISA \underline{E}xtension for mobile in-cache computing.
The proposed ISA provides multi-dimensional strided and random memory accesses, as well as efficient dimension-level masked execution. This flexible instruction set increases the Data Level Parallelism (DLP) exposed to the in-cache vector engine to multiple dimensions of the mobile kernels. A higher DLP allows utilizing all 8192 in-cache SIMD compute lanes with few vector instructions.

To realize the proposed \gpic{} ISA, we design a compute-capable cache architecture that targets small private L2 caches for a tight integration with the core. The proposed architecture leverages fine-grain communication between the core and L2 cache to interleave scalar and vector instructions between the scalar core and the in-cache vector engine.  Hence, \gpic{} provides an alternative way of scaling vector units for processing fine-grain data-parallel mobile kernels.

This paper offers the following contributions:

\begin{itemize}

    \item We propose a long-vector ISA extension that encodes parallelism across multiple dimensions using multi-dimensional strided and random memory accesses and efficient dimension-level masking.
    Compared to RISC-V RVV~\cite{riscv}, this instruction set improves the in-cache vector engine utilization from 23\% to 60\% and enhances the performance by 3.8$\times$, on average.

    \item We design cache architecture for multi-dimensional vector ISA. We analyze and evaluate different In-SRAM computing schemes for the proposed architecture, including bit-serial~\cite{neuralcache}, bit-parallel~\cite{bsbppincache}, bit-hybrid~\cite{eve}, and associative computing~\cite{cape}.

    
    \item We validate the effectiveness of our solutions on twelve real-world mobile libraries~\cite{swan}.
    An end-to-end system design with a long-vector bit-serial in-cache computing model improves the performance and energy of commercial vector CPUs by 2.9$\times$ and 8.8$\times$.

    \item We compare \gpic{} in-cache vector processing to a mobile GPU. Our measurements show that domain-specific accelerators (GPUs and DSPs) suffer from the kernel launch and data copy overhead.
    This is crucial for the fine-grain data parallelism, which benefits from a long-vector engine tightly integrated with the core. 

    \item We analyze and find that coarse-grain Single-Instruction Multiple-Thread (SIMT) models proposed for server class in-cache computing (e.g. Duality Cache~\cite{dualitycache}) are inefficient for mobile kernels due to the high compute latency and numerous register spills and fills.
    
\end{itemize}

\vspace{-1mm}

\section{Background}~\label{sec:background}

\vspace{-6mm}

\subsection{Long Vector Processing}

\textbf{Long-vector ISA Extensions.}
Table~\ref{tab:ISACMP} compares the state-of-the-art long-vector ISA extensions, \textit{i.e.,} RISC-V vector extension (RVV)~\cite{riscv}, Arm Scalable Vector Extension (SVE)~\cite{armsve}, and NEC~\cite{NEC_SX}.
SVE and NEC support up to 2K bit and 16K bit vector lengths.
Using SVE and NEC for the in-cache long-vector engine requires numerous instructions to target all 8K 32 bit SIMD lanes, increasing the core-cache communication.

RVV supports 1D strided vector accesses.
NEC provides 2D strided loads and stores restricted to a constant 16$\times$16 input matrix size.
An efficient long-vector ISA extension tailored for the small 1D DLP of mobile applications requires \textit{multi-dimensional} and \textit{flexible} strided memory accesses.

RVV offers random memory accesses from a base address and predetermined offsets stored in an index vector register.
SVE supports both the random base and random offset accesses.
These instructions are not efficient for capturing the random memory accesses in mobile kernels, which often have multiple unique base addresses in the highest dimension, followed by strided patterns for lower dimensions. 

Finally, all three long-vector ISA extensions support predicated execution for irregular DLP patterns. However, generating predicate masks with scalar or long-latency in-cache vector instructions for 8K SIMD lanes is expensive.
We introduce a coarse-grain dimension-level masking approach for the multidimensional access patterns.

\begin{table}[t]
  \centering
  \caption{Vector ISA Extension Comparison}
  \vspace{-2mm}
  \scalebox{0.95}{
  \setlength{\tabcolsep}{2.5pt} 
      \renewcommand{\arraystretch}{1.1}
  \begin{tabular}{|c|c|c|c|c|}
    \hline 
    \multirow{2}{*}{\textbf{Feature}} & \textbf{\gpic{}} & \textbf{RISC-V} & \textbf{Arm} & \multirow{2}{*}{\textbf{NEC~\cite{NEC_SX}}} \\
    & (This Work) & \textbf{RVV~\cite{riscv}} & \textbf{SVE~\cite{armsve}} & \\
    \hline \hline
    \multirow{2}{*}{\shortstack{Max Vector \\ Length}} & \multirow{2}{*}{\shortstack{infinite}} & \multirow{2}{*}{\shortstack{infinite}} & \multirow{2}{*}{\shortstack{2048 bits}} & \multirow{2}{*}{\shortstack{16384 bits}} \\
    & & & & \\
    \hline
    \multirow{2}{*}{\shortstack{Strided \\ Access}} & Flexible & Flexible & \multirow{2}{*}{-} & Constant \\
    & 4D & 1D & & 2D \\
    \hline
    \multirow{2}{*}{\shortstack{Random \\ Access}} & Random Base + & Random & Random Base / & \multirow{2}{*}{-} \\
    & Strided Offset & Offset & Random Offset & \\
    \hline
    \multirow{2}{*}{\shortstack{Masked \\ Execution}} & Predicate / & \multirow{2}{*}{Predicate} & \multirow{2}{*}{Predicate} & \multirow{2}{*}{Predicate} \\
    & Dimension-Level & & & \\
    \hline
   \end{tabular}
   }
  \label{tab:ISACMP}
  \vspace{-5mm}
\end{table}

\textbf{Long-vector Engines} are designed and fabricated by both the industry~\cite{alibaba_Xuantie, andes_NX27V, SiFive_P270, SiFive_X280, NEC_SX, Fugaku_A64FX} and academia~\cite{Hwacha, Ara, Arrow, vicuna, vitruvius, vitruvius_plus}.
These units are mostly employed by the supercomputers with abundant area and power resources.
For example, Vitruvius+~\cite{vitruvius_plus} is an RVV-based coprocessor which has 16K bit vector registers. The vector units though have 8 dual-precision or 16 single-precision SIMD lanes. Each instruction is scheduled on the SIMD lanes in 64 cycles in a vector manner. This coprocessor requires an area of $1.3mm^2$ at a 22nm technology ($\sim0.4mm^2$ at 7nm) which is significant compared to the evaluated Arm Cortex-A76 core ($\sim1.2mm^2$ at 7nm), including two 128-bit Neon ASIMD units and private L1 and L2 caches. 

Due to the high area overhead of long-vector engines, all Arm application processors employ 128-bit ASIMD units since the introduction of Arm Neon technology.
\gpic{} utilizes the untapped private L2 cache resources and re-purposes them to 8K$\times$32 bits in-cache bit-serial SIMD lanes private to each core with 3.6\% area overhead to the core.

\subsection{In-Cache Computing}
\label{subsec:insram_computing}

\begin{figure}[t]
	\centering
    \includegraphics[width=.48\textwidth]{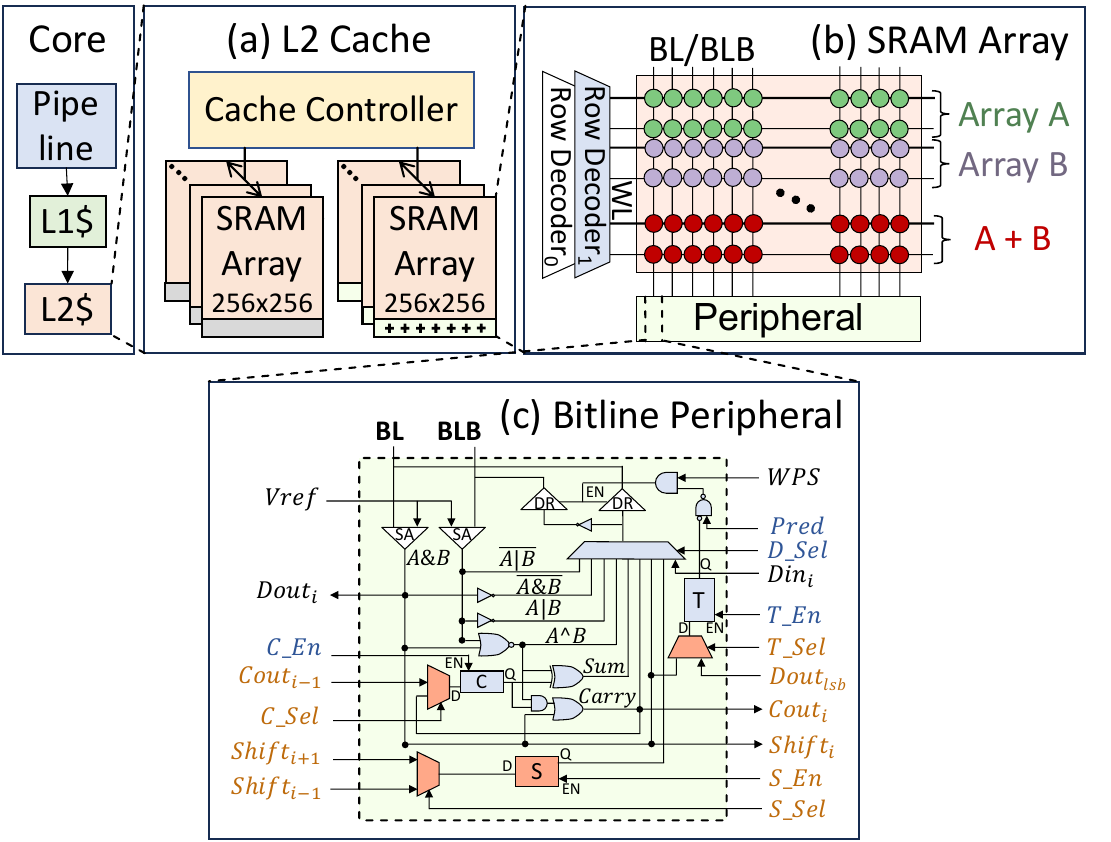}
    \vspace{-1mm}
    \caption{
    (a) Mobile core with in-cache computing enabled for half of the L2 cache.
    (b) In-SRAM computing activates two word-lines of an SRAM array using an extra row decoder.
    (c) Bit-Serial (blue), Bit-Hybrid, and Bit-Parallel (blue + orange) modifications to the bitline peripheral.}
    \label{fig:in_sram}
    \vspace{-6mm}
\end{figure}

In-Cache computing~\cite{sramlogic, energyefficient, computecaches, neuralcache, mlimp} repurposes cache SRAM arrays to SIMD units.
Figure~\ref{fig:in_sram}(a) shows a mobile core with half of the L2 cache repurposed for in-cache computing.
Neural Cache~\cite{neuralcache}, which we build upon, proposes data elements are vertically aligned to the SRAM bitlines.
SRAM peripherals compute bit-serial operations in-parallel for all data elements, as shown in Figure~\ref{fig:in_sram}(b).
Therefore, a SRAM array with 256 bitlines provides 256 SIMD lanes, and a 256KB L2 cache slice with 32 SRAM arrays can embody an 8192-lane bit-serial vector engine. In this section, we describe the architecture of prior In-SRAM computing schemes:




(a) \textbf{Bit-Serial Computing (BS)~\cite{neuralcache}} vertically aligns the array elements in the bitlines.
By activating two word-lines of an SRAM array, Sense Amplifiers ($SA$) at the peripheral logic calculate the logical AND and NOR of a bit-slice across all elements of two input arrays.
Extra logic is employed (blue components of Figure~\ref{fig:in_sram}(c)) to compute the NAND, OR, and XOR of bit-slices.
An $n$-bit integer addition is performed in $n$ cycles using additional logic to compute the sum and carry-out. A Carry latch ($C$) stores the carry-out in the bitline peripheral as the carry-in of the next bit.
A two's complement subtraction is computed in $2n$ cycles by negating $B$ and adding $A$ and $\overline{B}$ using the carry-in ($C$) set to 1.
Integer multiplication requires $n^2+5n$ cycles.
First, each $i^{th}$ bit of the multiplicand is stored in the Tag latch ($T$) as an enabling signal to the bitline drivers.
Next, the multiplier is conditionally added to the results, starting from the $i^{th}$ bit.
Constant shift operation is supported in $n$ cycles by reading input bit-slices and writing to bit-slices of the shifted result, offset by the defined shift value.
Variable shift needs $\mathcal{O}(n \log n)$ cycles~\cite{dualitycache} to read each $i^{th}$ bit of $B$ to the $T$ latch and perform conditional constant shift with $2^i$ bit value. Duality Cache~\cite{dualitycache} extends these  integer operations to floating-point arithmetic operations.

\ignore{
\ignore{~\cite{bitserialincache}.
Compute Caches~\cite{computecaches} and VRAM~\cite{bsbppincache} employ extra logic (blue components of Figure~\ref{fig:in_sram}) to perform NAND, OR, and XOR operations.
An \textit{8:1} multiplexer, implemented using pass transistors~\cite{bsbppincache}, selects the required logic, and bitline drivers ($DR$) store the output back to the SRAM array.}

For floating-point addition, Duality Cache subtracts the exponents, exploits the variable shift operation to normalize the mantissa of the operands, and adds the normalized mantissa.
For multiplication, Duality Cache adds the exponents, multiplies the mantissa, and normalizes the exponents and mantissa using subtraction and constant shifts.
}

(b) \textbf{Bit-Parallel~\cite{bsbppincache}~(BP) and Bit-Hybrid~\cite{eve}~(BH).}
BS maximizes parallelism to 8K SIMD lanes but has high arithmetic latency.
VRAM~\cite{bsbppincache} introduces BP, aligning $n$-bit data horizontally in a word-line.
This reduces parallelism to $8K/n$ SIMD lanes yet improves latency by a factor of $n$.
EVE~\cite{eve} introduces BH by merging BS and BP schemes to balance the latency and parallelism.
BH splits $n$-bit data into $p$-bit segments were segments are computed using BP, and combined using BS.
BP and BH computing require communication across bitline peripherals which incurs area and frequency overheads~\cite{bsbppincache}.
Figure~\ref{fig:in_sram}(c) shows the additional logic for the inter-bitline communication in orange.
EVE~\cite{eve} employs Manchester carry chain and routes carry-out to the next bitline for the full addition inside a segment while the $C$ flip-flop is used for the addition between segments. 



(c) \textbf{Associative Computing (AC)~\cite{cape}.}
In contrast to the prior schemes, AC does not employ extra logic in the bitline peripheral for logical operations.
Instead, AC employs the search and update operations of the Binary Content Addressable Memory (BCAM) structures~\cite{bcam}.
To perform any logical operation, AC executes a search and update for each row of the truth table.
Using AC, CAPE~\cite{cape} horizontally aligns a bit-slice of all array elements in a word-line.
Since CAPE distributes different bit-slices in different SRAM arrays, it executes a bit-wise operation in parallel for all bits in $\mathcal{O}(1)$ cycles.
Prior work~\cite{cape_db} exploits the low latency of these bit-wise logical operations to accelerate database workloads.
However, the search and update operations are performed sequentially for different bits of full addition, since it requires carry propagation between bit slices.
An $n$-bit integer subtraction/addition requires $8n+2$ cycles.
Because other arithmetic operations are decomposed to addition and subtraction, they incur high operation latency as well. 

In \textbf{summary}, all In-SRAM computing schemes convert the cache SRAM arrays to a high-throughput long-vector engine.
Therefore, \gpic{} is required to utilize the multi-dimensional DLP of mobile kernels for the numerous in-cache SIMD lanes.
While Section~\ref{subsec:isa_micro_perf} evaluates this claim, we use the bit-serial computing to explain our contributions throughout this paper.
\section{Instruction Set Architecture}

We draw on the insights from analyzing Swan~\cite{swan}, a state-of-the-art benchmark suite of data-parallel applications for mobile processors. We find that multidimensional data structures are prevalent in mobile applications. These insights inform our design choices, allowing us to effectively exploit mobile workload characteristics in the proposed \gpic{} extension.




\subsection{Design Goals}

Given the high throughput of the in-cache computing engine, the \textit{key ISA design challenge} is to achieve high compute resource utilization by developing a \textit{compact ISA} that expresses vector computation in few instructions. 
An ISA that generates numerous dynamic instructions leads to instruction issue bottlenecks and overflows the micro-architecture modules.
Consequently, the scalar core cannot issue sufficient instructions to keep in-cache vector units busy.

In addition, prior in-cache ISA extensions~\cite{blade, instructioncode, computecaches} expose the cache geometry and data layout to programmers, requiring them to load and align data elements along specific SRAM bitlines to ensure operand locality.
This approach increases programming complexity, as programmers must deal with cache bank, set, way, word-line, and bitline coordinates, resulting in binaries incompatible with other cache architectures.
In contrast, we aim to abstract away these complexities by presenting in-cache computing as a long-vector processing engine.
Programmers can use in-cache vector instructions and registers similar to other long-vector extensions, while the underlying architecture transparently maps long-vector registers to SRAM bitlines and word-line coordinates.

\subsection{Physical Register Abstraction.}
\label{subsec:physical_registers}

\begin{figure}[t]
    \centering
    \includegraphics[width=.49\textwidth, trim={0.1cm 0.5cm 0.1cm 0.1cm}, clip]{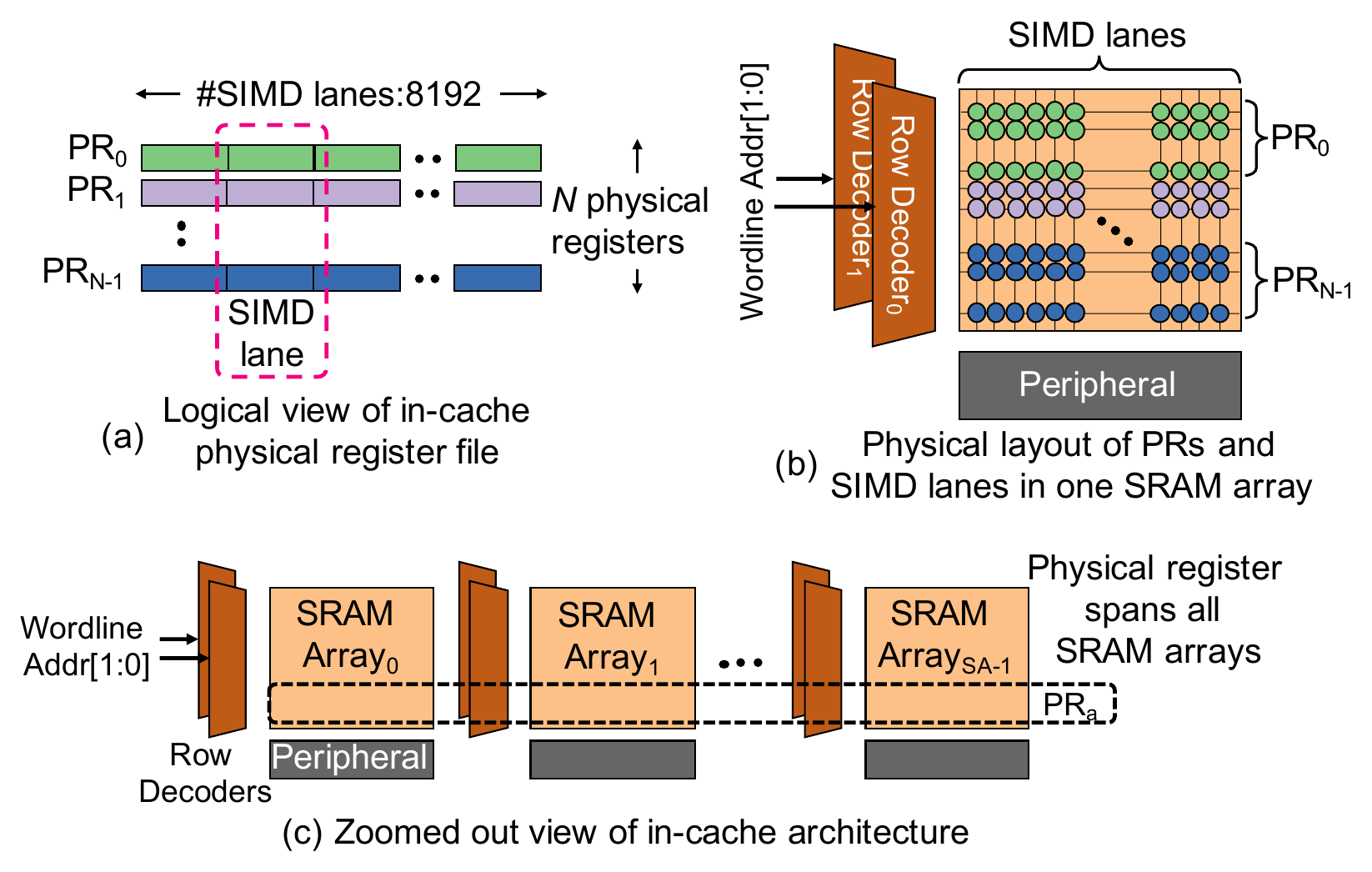}
    \vspace{-5mm}
    \caption{(a) \gpic{} operates on \textit{N} long-vector in-cache registers. (b) In-cache data elements and SIMD lanes use the vertical data layout of bit-lines. (c) An in-cache physical register spans all compute-capable SRAM arrays.}
    \label{fig:register_organization}
    \vspace{-5mm}
\end{figure}

Figure~\ref{fig:register_organization} shows the data layout of vector registers in the cache:
(a) \gpic{} supports $N$ Physical Registers ($PR_{0}$ to $PR_{N-1}$).
(b) Internally, data elements of a physical register are transposed and placed vertically in a bitline to support bit-serial In-SRAM operations.
Thus, an SRAM array with 256 bitlines provides 256 data elements for a PR.
(c) Each PR spans across all SRAM arrays.
Therefore, an in-cache long-vector PR contains 8K data elements using 32 SRAM arrays.

We observe that the mobile data-parallel kernels of the Swan benchmark suite~\cite{swan} operate on multi-dimensional data structures using nested loops and utilize at most four dimensions.
Further, mobile data-parallel kernels often operate on small input data sizes, so one dimension is insufficient to harness the throughput of in-cache vector engines.
Thus, rather than addressing each of 8K SIMD lanes in a linear address space (\textit{i.e.,} $PR_{i} [s]$), \gpic{} ISA treats physical registers as multi-dimensional logical registers (\textit{i.e.,} $PR_{i} [w][z][y][x]$). 

\gpic{} cache controller (Section~\ref{sec:design}) maintains the vector engine configurations in specific \textit{Control Registers (CR)}.
Programmers select the dimension count and lengths using config instructions that set the CRs.
Logical register indices are flattened-out to the 8K SIMD lanes and mapped to a \textit{\gpic{} physical register} by the \gpic{} controller.
Many kernels use the same memory access pattern for multiple input batches.
Therefore, the overhead of configuring CRs is amortized over multiple vector executions.

The physical vector register file in the conventional vector architectures has a fixed number of registers with a constant width, such as 40 16K bit registers in the Vitruvius+~\cite{vitruvius_plus} architecture.
Vector ISA extensions, like RVV~\cite{riscv}, offer a constant number of registers but vary the vector length based on the data precision.
In contrast, the bit-serial layout of the In-SRAM computing offers a constant vector length (8K bitlines) but allows a variable number of registers based on the data widths.
While a fixed vector length regardless of the data precision simplifies the vectorization, a variable number of registers complicates register allocation (Section~\ref{subsec:compiler}).

\gpic{} compute instructions perform the same operation on all SIMD lanes of the physical register, yet loads and stores to the physical registers can target different memory locations.
As we discuss next, the multi-dimensional logical vector register abstraction allows offloading versatile and multi-dimensional memory access patterns.

\subsection{Strided Memory Access with Possible Replication}
\label{sma}

The memory access of data-parallel kernels follows a regular pattern: nested loops process multiple data dimensions, and the base memory address of each iteration differs with a constant stride. Therefore, we propose multi-dimensional strided memory access instructions as shown in Algorithm 1.


\begin{algorithm}[ht]
    \small
    \caption{Strided Multi-dimensional Vector Load}
    \begin{algorithmic}[1]
        \renewcommand{\algorithmicrequire}{\textbf{Input:}}
        \renewcommand{\algorithmicensure}{\textbf{Output:}}
        \REQUIRE $Dim_{[3:0]}.Length:$ CRs, $Base$ Address, and $S_{[3:0]}:$ Strides
        \FOR {$w$ in $[0 : Dim_3.Length)$}
            \FOR {$z$ in $[0 : Dim_2.Length)$}
                \FOR {$y$ in $[0 : Dim_1.Length)$}
                    \FOR {$x$ in $[0 : Dim_0.Length)$}
                        \STATE $PR_i[w][z][y][x] = MEM [Base + w \times S_3 + z \times S_2 +  y \times S_1 + x \times S_0]$
                    \ENDFOR
                \ENDFOR
            \ENDFOR
        \ENDFOR
    \end{algorithmic} 
    \label{alg:stridedmemoryaccess}
\end{algorithm}

\begin{figure}[h]
    \centering
    \vspace{-2mm}
    \includegraphics[width=.45\textwidth]{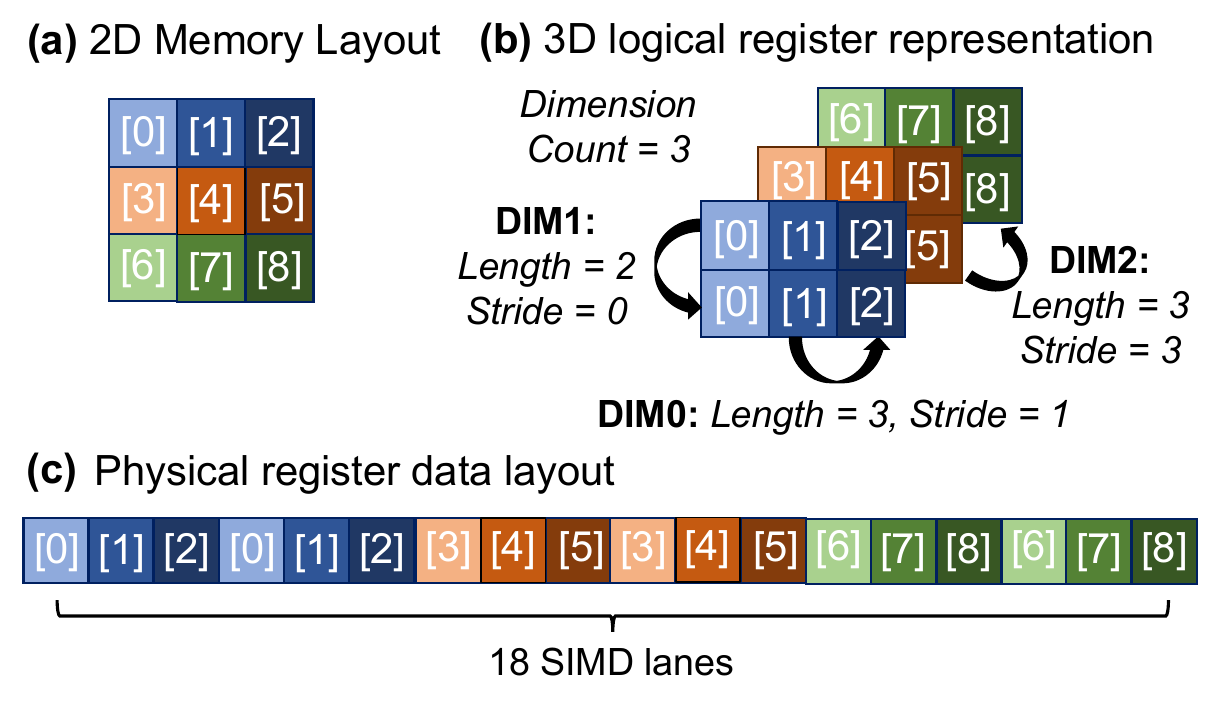}
    \vspace{-4mm}
    \caption{Strided memory access example of \textit{Intrapicture Prediction} kernel: loading from (a) 2D memory layout to (b) 3D logical registers, mapped to (c) the SIMD lanes of flattened-out physical registers by \gpic{} controller.}
    \label{fig:stridedmemaccess}
    \vspace{-1mm}
\end{figure}

As an example, a typical pattern employed in the \textit{Intrapicture Prediction} kernel of image and video coding~\cite{hevc} is shown in~Figure~\ref{fig:stridedmemaccess}. This kernel needs a 3D strided vector load with $S_0=1$, $S_1=0$, and $S_2=3$.
The input data is located in a 2D data structure but is loaded into a 3D logical register for the actual computation. 
The strided vector load instruction encodes data loading from 2D to 3D along with the replication in a single instruction.
The stride $S_1=0$ facilitates the replication of each element in an input data row to an entire column in the 3D logical register. The figure further shows a flattened-out data mapping from a 3D logical register representation to different SIMD lanes of a physical register. 
\setlength{\textfloatsep}{0pt}

1D vector ISA extensions like RVV~\cite{riscv} would employ 6 strided load instructions to access each row of the input data (\textit{e.g.}, $[0][1][2]$) separately to a 3-element segment of the vector register, while masking off the other segments. Further scalar instructions are needed to compute the mask and 6 instructions to load them. \gpic{} encodes this pattern in a single instruction without needing any masks.

Each stride value ($S_i$) takes up to 16 instruction bits.
Encoding multiple stride values for different dimensions increases the instruction width.
However, our analysis shows that stride values of 0 (for replication) and 1 (for sequential access) are frequently used.
Thus, instead of a 16-bit absolute stride value, we encode a 2-bit \textit{stride mode} for each dimension (8 bits for four dimensions).
Modes $0$ and $1$ encode frequently-used stride values of $0$ and $1$.
$Mode~2$ is designed for the sequential accesses, where the stride in the higher dimensions differs.
For example, when loading from a row-wise 2D matrix, the stride value of the row dimension equals to the number of columns.
Therefore, using mode 2 for the $i^{th}$ dimension provides a stride of $S_i=S_{i-1}\times Dim_{i-1}.Length$.
If the desired stride value is neither of the mentioned modes, we provision a load and a store stride CR for each dimension in \gpic{} controller, which is set by the config instructions.
$Mode~3$ uses these CR values.

\subsection{Random Memory Access with Striding and Replication}
\label{rma}
\begin{figure}[t]
	\centering
    \includegraphics[width=.48\textwidth]{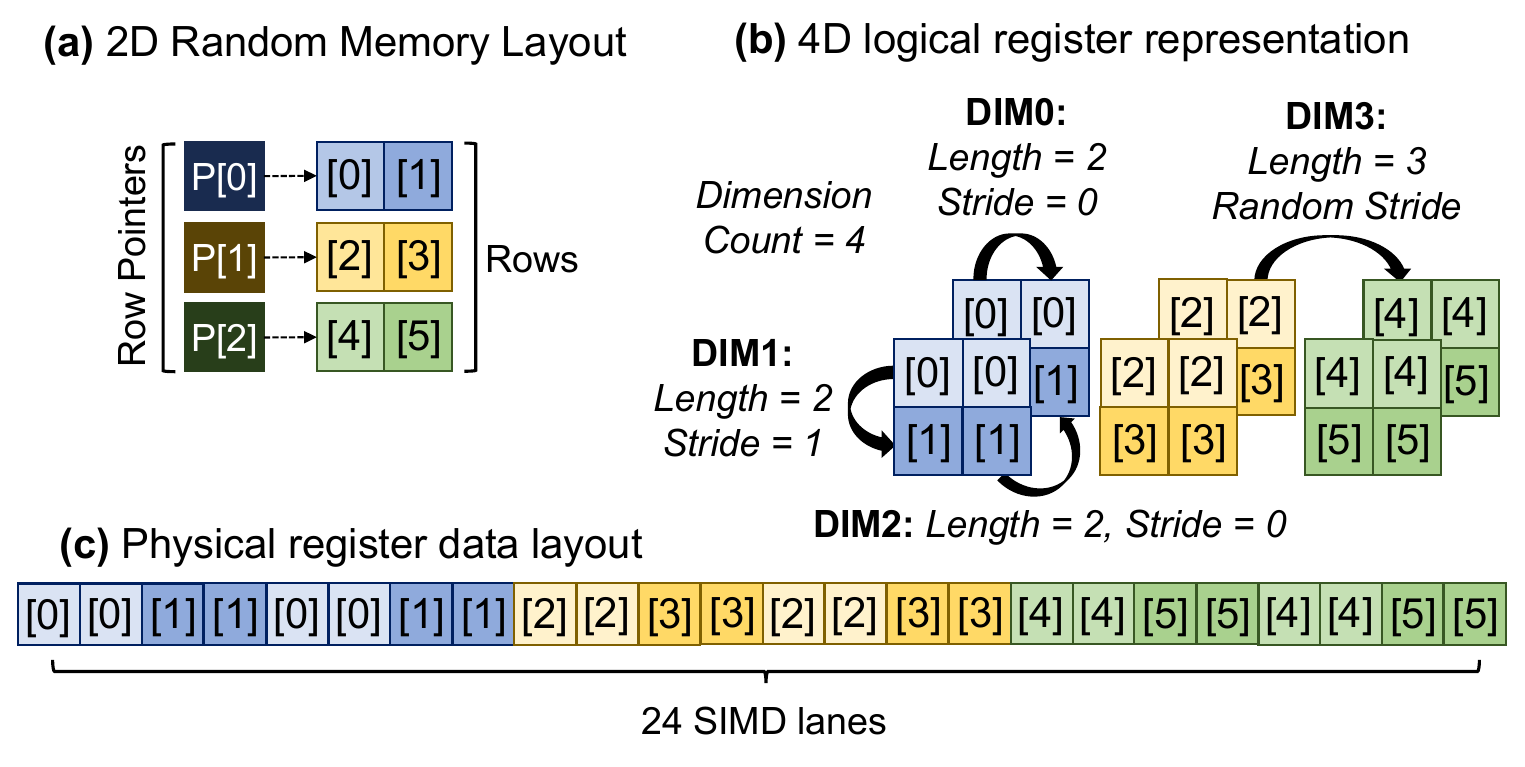}
    \vspace{-3mm}
    \caption{Random memory access of \textit{h2v2 Upsample} kernel: loading from (a) random row pointers to (b) 4D logical registers. (c) shows the SIMD lanes of the flattened-out physical registers.}
    \label{fig:randommemaccess}
\end{figure}

We observe that in some mobile data-parallel kernels, the outer-most loop (highest dimension) requires unique base addresses, while the inner loops follow the stride pattern.
Figure~\ref{fig:randommemaccess} demonstrates this concept with an example of \textit{libjpeg} library~\cite{libjpeg}.
\textit{libjpeg} allocates rows in separate memory locations. \textit{h2v2 Upsample} kernel replicates each pixel horizontally and vertically into 4 pixels.
While the address in the 4th dimension is random (row pointers), other dimensions follow the stride values of $0$ (replicate vertically), $1$ (row pixels), and $0$ (replicate horizontally).

The random memory access instruction of \gpic{} encodes the address of the input pointer array.
\gpic{} controller fetches the random base addresses from this array, and employs stride values for the inner dimensions as shown in Algorithm~\ref{eq:randommemoryaccess}.

\vspace{-4mm}
\begin{equation} \label{eq:randommemoryaccess}
    PR_i[w][z][y][x] = MEM[Base_w + z \times S_3 +  y \times S_2 + x \times S_1]
\end{equation}

RVV~\cite{riscv} supports a single base address with random offsets located in a vector register. Implementing the above memory access pattern with these instructions requires 3 masked random memory accesses per base address, scalar instructions to compute mask, and 6 extra  memory accesses to load the mask and offset vector registers. In contrast, \gpic{} encodes this pattern in a single random memory access without scalar instruction overhead.

\ignore{
On the other hand, the random memory accesses of 1D vector ISA extensions such as RVV~\cite{riscv} take a single base address and use random offsets located in a vector register.
Using these instructions to implement the above memory access pattern requires 3 masked random memory accesses to load from each base address separately.
The mask and offset vector registers are computed by the scalar core, stored to the memory, and loaded to the vector registers using 6 extra memory accesses.
\gpic{} encodes this pattern in a single random memory access without any scalar instruction overhead.
}

\subsection{Efficient Masked Execution} 
\begin{figure}[t]
    \centering
    \includegraphics[width=.38\textwidth, trim={4mm 3mm 4mm 3mm}, clip]{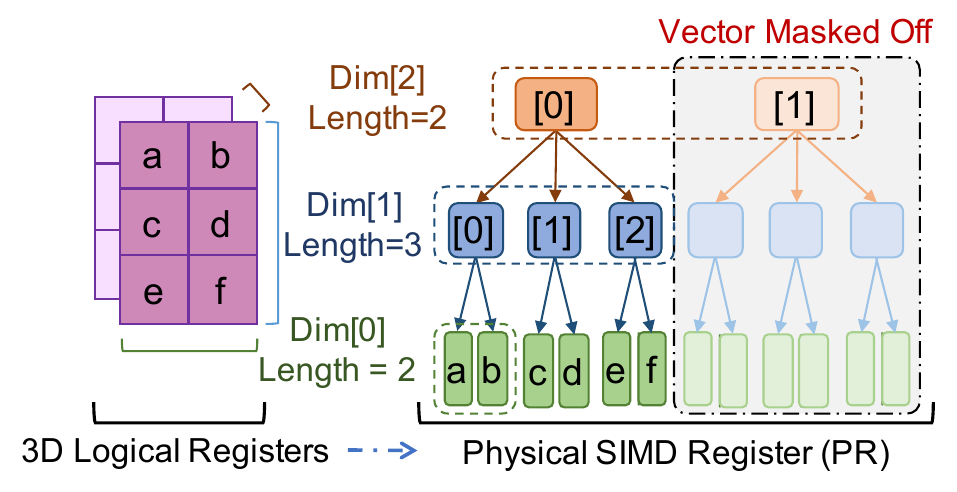}
    \caption{\gpic{} Controller maps multi-dimensional \textit{logical registers} to 1D \textit{Physical SIMD Registers}.
    Efficient dimension-level masked execution masks off leaves under a node in the highest dimension of the tree (iterations of the outer-most loop).}
    \label{fig:MEGAmask}
\end{figure}

\gpic{} supports masked execution using two types of instructions:
\textit{(a) Conventional Predictated Execution:}
\gpic{} provides compare operations whose result are stored in the Tag Latch (T) to mask off each SIMD bit-line independently 
(leaves of dimension tree in Figure~\ref{fig:MEGAmask}).
(b) \textit{Dimension-Level Mask Operations}: We observe that most of the data-parallel kernels require \textit{coarse-grain} masked execution where SIMD lanes corresponding to the outermost loop are masked off (leaves under the orange node $[1]$).
We keep a mask CR that contains one bit for each element of the highest dimension.
We set the maximum length of the highest dimension to 256 to limit the mask CR size.
Mask values are calculated by the scalar instructions and set in the \gpic{} controller using vector mask instructions, that mask on/off SIMD lanes of a specific element of the highest dimension.
Using conventional predicated execution to implement this semantic requires computing the mask value of each element in the scalar core, storing them to the memory, and loading them to a vector mask register.
Dimension-level mask operations are efficient as they do not use In-SRAM physical registers or scalar instructions.

\subsection{Programming Model}\label{subsec:programming}

\gpic{} supports 8/16/32/64-bit un/signed integer and 16/32-bit floating-point data types denoted by \textit{b/w/dw/qw} and \textit{hf/h} suffixes.
29 operations are implemented for each data type, which are classified in Table~\ref{tab:ISA}.
\textit{Config} operations set the CRs of the \gpic{} controller and are not executed on SRAM arrays.
\textit{Move} operations convert and copy registers.
\textit{Load} and \textit{Store} operations provide both strided and random vector memory accesses.
\textit{Arithmetic} operations are used to set the value of a register based on source registers and immediate values.

\begin{table}[b]
\scriptsize
\renewcommand\arraystretch{1.05}
    \centering
    \caption{\gpic{} Instructions}
    \label{tab:ISA}    
    \vspace{-2mm}
    \scalebox{1}{
        \begin{tabular}{|c|c|c|c|}
        \hline
        \textbf{Type} & \textbf{Operation} & \textbf{Assembly} & \textbf{BS Latency*} \\
        \hline
        \multirow{4}{*}{Config} & set dim count & \textit{vsetdimc rs} & \multirow{4}{*}{$-$} \\
        \cline{2-3}
        & set dim length & \textit{vsetdiml rs1 rs2} & \\
        \cline{2-3}
        & (un)set mask & \textit{v(un)setmask rs} & \\
        \cline{2-3}
        & set kernel width & \textit{vsetwidth imm8} & \\
        \hline
        \multirow{2}{*}{Move} & convert & \textit{vcvt vd vs} & \multirow{2}{*}{$n$} \\
        \cline{2-3}
        & copy & \textit{vcpy vd vs} & \\
        \hline
        & strided load & \textit{vsld vd rs1 rs2} & \multirow{4}{*}{$-$} \\
        \cline{2-3}
        Memory & random load & \textit{vrld vd rs1 rs2} & \\
        \cline{2-3} 
        Access & strided store & \textit{vsst vs rs1 rs2} & \\
        \cline{2-3}
        & random store & \textit{vrst vs rs1 rs2} & \\
        \hline
        \multirow{10}{*}{Arithmetic} & set duplicate & \textit{vsetdup vd rs} & $n$ \\
        \cline{2-4}
        & shift immediate** & \textit{vshi vd vs rs} & $n$ \\
        \cline{2-4}
        & rotate immediate** & \textit{vroti vd vs rs} & $n$ \\
        \cline{2-4}
        & shift register** & \textit{vshr vd vs1 vs2} & $nlog(n)$ \\
        \cline{2-4}
        & addition & \textit{vadd vd vs1 vs2} & $n$ \\
        \cline{2-4}
        & subtraction & \textit{vsub vd vs1 vs2} & $2n$ \\
        \cline{2-4}
        & multiplication & \textit{vmul vd vs1 vs2} & $n^2+5n$ \\
        \cline{2-4}
        & min/max & \textit{vmin/max vd vs1 vs2} & $2n$ \\
        \cline{2-4}
        & xor & \textit{vxor vd vs1 vs2} & $n$ \\
        \cline{2-4}
        & comparison & \textit{vgt(e)/lt(e)/(n)eq vs1 vs2} & $n$ \\
        \hline
        \multicolumn{4}{l}{*Compute latency of bit-serial signed integer operations based on precision (n).} \\
        \multicolumn{4}{l}{**Shift and rotate operations include \textit{right (r)} or \textit{left (l)} variants.} \\
        \end{tabular}
    }
\end{table}

To use \gpic{} data types and instructions in the C/C++ program, we developed a library of supported variables and intrinsics.
\gpic{} variables are declared as \textit{\_\_mdv} as an acronym for Multi-Dimensional Variable, concatenated by the data type suffixes.
Programmers include this library in the code and develop their kernels using \gpic{} with a comparable level of effort to that required for other long-vector architectures (\textit{e.g.}, RVV~\cite{riscv} and SVE~\cite{armsve}).
We developed an in-house compiler infrastructure and applied the following compiler optimizations on the data-parallel \gpic{} kernels.
Executable binaries are decoded in the front-end modules of the core, the scalar back-end pipelines execute scalar instructions, and long-vector in-cache instructions are issued to the \gpic{} Controller.

\subsection{Compiler}~\label{subsec:compiler}
Compiler support of \gpic{} is similar to long-vector architectures with the following considerations:

\textbf{Register Count.}
While conventional architectures define a constant number of PRs, \gpic{} PRs are limited by the number of word-lines (256) and the width of live PRs (Figure~\ref{fig:register_organization}(b)).
To simplify register allocation, we employ a single data width for a kernel.
Using Liveness Analysis~\cite{livenessanalysis}, compiler detects the widest PR width of a kernel and injects a config instruction (Table~\ref{tab:ISA}) to set the PR width as a control register.
\gpic{} controller maps the PRs to the word-lines using this CR.

\textbf{Register Allocation and Instruction Scheduling}
are two Code Generation optimizations that improve register spills and ILP.
We focus on vector register pressure as it is critical to \gpic{} due to the limited number of PRs and the high spill cost of wide in-cache vector registers.
We use \textit{list-hybrid} instruction scheduler~\cite{llvmcodebook}, a bottom-up List Scheduling algorithm that tries to keep the live registers less than the target machine's PR count and shorten register live ranges by scheduling instructions that do not generate register spills close to their use.
Moreover, we employ \textit{Greedy Register Allocation}~\cite{greedyra}, which minimizes register spills by extensively splitting live ranges, reducing register interference.

\section{Common Data-Parallel Patterns}\label{sec:patters}

To illustrate the expressibility of the \gpic{} ISA, we show the implementation of common data-parallel patterns of the evaluated mobile kernels.
The instructions are listed in Table~\ref{tab:ISA} concatenated with the data type suffixes.

\textbf{Matrix transposition} is employed in the Machine Learning (\textit{XNNPACK}~\cite{xnnpack}), image/video codecs (\textit{Kvazaar}~\cite{kvazaar} and \textit{libjpeg}~\cite{libjpeg}), and cryptography (\textit{boringssl}~\cite{boringssl}) libraries.
The following code snippet shows the transposition of an $M \times N$ matrix using \gpic{}.
For simplicity, in this example, we assume both dimensions are smaller than the number of SIMD lanes.
Any kernel starts with configuring the dimension count, lengths (\underline{L}ine 3), and stride CRs (L5).
The main loop of the transpose kernel loads the input matrix columns vertically using the stride values of \texttt{N} and \texttt{1} (L9) and stores the output matrix rows horizontally with the stride values of \texttt{1} and \texttt{N} (L12).
To transpose a $512\times49$ matrix (one of MobileNet-V1~\cite{mobilenet} kernels), the \gpic{} implementation requires only 4 iterations, while the 1D implementation needs 49 iterations to load/store input columns separately.

\begin{lstlisting}[style=CStyle]
void transpose(int input[M][N], int output[N][M]){
  // 2D: M output columns (DIM0),8192/M rows (DIM1)
  vsetdimc(2); vsetdiml(0, M); vsetdiml(0, 8192/M);
  // load stride(DIM0) = N, store stride(DIM1) = N
  vsetldstr(0, N); vsetststr(1, N);
  for (int i = 0; i < N; i += 8192/M) {
    // LD 8192/M columns: stride(DIM0) = MODE3 (N)
    //                    stride(DIM1) = MODE1 (1)
    __mdvdw in_val = vlds_dw(&input[i], 3, 1);
    // ST 8192/M rows: stride(DIM0) = MODE1 (1)
    //                 stride(DIM1) = MODE3 (N)
    vsts_dw(&output[i*M], in_val, 1, 3); 
  } 
}  
\end{lstlisting}

\textbf{Reduction} is employed in the Network (\textit{Checksum}~\cite{optroutines}), audio/image processing (\textit{libwebp}~\cite{libwebp} and \textit{webaudio}~\cite{webaudio}), and compression (\textit{zlib}~\cite{zlib}) libraries.
The reduction of an input array to 8K elements (number of SIMD lanes) is simply implemented using 1D vector loads and additions.
The reduction beyond 8K elements repeatedly splits the vector register into two halves and then reduces to half of the elements.
Instead of supporting arbitrary data movements between SRAM arrays and bit-lines to align the half registers, we mask off the first half of the register using dimension-level mask operations (L6) and store the second half of the vector register to the memory using 2D vector stores (L8).
Next, we set the vector length to half (L10), load the second half into a temporary register (L11), and reduce the half vector registers (L13).
We repeat this vertical reduction step until we reduce the array to 256 elements (5 steps) after which the reduction proceeds in the CPU core, as in-cache computing is not beneficial after this point due to the significant arithmetic operation latency.

 
\begin{lstlisting}[style=CStyle]
__mdvdw vertical_reduction_step(__mdvdw a, int M) {
  int tmp_mem[M];
  // Split M SIMD lanes into 2 M/2-element halves
  vsetdimc(2); vsetdiml(1, 2); vsetdiml(0, M/2);
  // Mask off the first half (element 0)
  vunsetmask(0);
  // Store the second half to the temp memory 
  vsts_dw(tmp_mem, a, 1);
  // Load the second half into a register
  vsetdimc(1); vsetdiml(0, M/2);
  __mdvdw tmp_val = vsld_dw(&tmp_mem[M/2], 1);
  // Reduce the two M/2-element registers
  return vadd_dw(a, tmp_val);
}
\end{lstlisting}

\textbf{Irregular accesses} require random vector loads and stores. We observe this pattern in two scenarios:
(a) Sparse computation: for example, in the \textit{SPMM} kernel of the \textit{XNNPACK}~\cite{xnnpack} library, the input matrix is sparse and compressed in a CSR format.
Core computes the non-zero input cell addresses based on the row-pointers, and \gpic{} randomly loads and replicates them horizontally.
Similarly, core computes the weight row addresses corresponding to non-zero input cells, and \gpic{} randomly loads them sequentially.
(b) Data pointers: the image processing (\textit{libjpeg}~\cite{libjpeg}, \textit{libpng}~\cite{libpng}, and \textit{libwebp}~\cite{libwebp}) and graphics (\textit{Skia}~\cite{skia}) libraries maintain the frame rows in random memory locations.
To operate on multiple rows simultaneously, we need to load or store different rows from random base addresses.
The following code snippet shows only the memory access pattern of the \textit{libjpeg}'s \textit{upsample} kernel.
The input pixels are loaded from random input row pointers and each pixel is replicated twice (L10).
The output results are stored to output pointers sequentially (L14).

\begin{lstlisting}[style=CStyle]
void upsample(char* input[M], char* output[2M]) {
  // 3D: replicate 2 pixels (DIM0)
  vsetdimc(3); vsetdiml(0, 2);
  // M columns (DIM1), 8192/(2M) rows (DIM2)
  vsetdiml(1, M); vsetdiml(2, 8192/(2M));
  for (int n = 0; n < N; n += 8192/(2M)) {
    // Random LD rows from input pointers
    // Load pixels in a row: stride(DIM1)= MODE1(1)
    // Rep. pixels twice: stride(DIM0)= MODE0(0)
    __mdvb rows_val = vrld_b(&input[n], 0, 1);
    // Random ST rows to output pointers
    // Store sequentially: stride(DIM0) = MODE1(1)
    //                     stride(DIM1) = MODE2(2)
    vrst_f(&output[n], rows_val, 1, 2);
  }
}
\end{lstlisting}

\textbf{Multidimensional replication} is prevalent in all data-parallel kernels.
For example, the row-wise \textit{GEMM} kernel of the \textit{XNNPACK}~\cite{xnnpack} library loads and replicates the input (L12) and weight (L15) elements horizontally (DIM0) and vertically (DIM1), respectively.
Setting the $i^{th}$ stride to zero replicates across the $i^{th}$ dimension as shown in the following code.

\begin{lstlisting}[style=CStyle]
void GEMM_Replication(float input[N][K],
    float weight[K][M], float output[N, M]) {
  // 2D: M output columns(DIM0), 8192/M rows(DIM1)
  vsetdimc(2); vsetdiml(0, M); vsetdiml(1, 8192/M);
  // load stride(DIM1)= K, store stride(DIM1)= M
  vsetldstr(1, K); vsetststr(1, M);
  for (int n = 0; n < N; n += 8192/M) {
    __mdvf acc_val = vsetdup_f(0);
    for (int k = 0; k < K; k++) {
      // LD an input column: stride(DIM1)= MODE3(K)
      // Rep. horizontally: stride(DIM0)= MODE0(0)
      __mdvf input_val = vsld_f(&input[n*K+k],0,3);
      // LD a weight row: stride(DIM0)= MODE1(1)
      // Rep. vertically: stride(DIM1)= MODE0(0)
      __mdvf weight_val= vsld_f(&weight[k*M],1,0);
      acc_v=vadd_f(acc_v,vmul_f(input_v,weight_v));
    }
    // ST sequentially
    vsst_f(&output[n*M], acc_val, 1, 3);
  }
}
\end{lstlisting}

\section{Design} \label{sec:design}

\subsection{Core Micro-architecture} \label{subsec:microarchitecture}

\gpic{} instructions are fetched, decoded, and pushed to both the Reorder-Buffer (ROB) and Load Store Queue (LSQ).
We do not support the out of order and speculative issue of \gpic{} instructions.
Hence, the LSQ prevents the reordering of scalar loads and \gpic{} stores.
Once committed, \gpic{} instructions are issued to the L2 cache at the head of the ROB, and retired from the LSQ and ROB.
Because none of the \gpic{} instructions write to the scalar registers (Table~\ref{tab:ISA}), no write back is required for these instructions.

\gpic{} stores, however, are pushed to the write buffer until completed and acknowledged by the \gpic{} controller.
The in-cache \gpic{} store execution takes many cycles, which stalls the younger scalar loads in the ROB.
To minimize unnecessary stalls, we provision an {Address Decoder} in the LSQ.
The address decoder maintains ${Dim Count}$, ${Dim_{[0:3]}.Length}$, and ${Dim_{[0:3]}.Stride}$ CRs that are updated upon the issue of the config instructions to the cache.
When a \gpic{} store is committed and pushed to the write buffer, the address decoder computes an \textit{address range} following Equation~\ref{eq:addrrange}.
Using this address range, the write buffer looks up for memory dependency between the scalar loads in the ROB and \gpic{} stores in the write buffer.

\vspace{-3mm}
\begin{equation} \label{eq:addrrange}
\begin{aligned}
    Range = Base + \sum_{i=0}^{3} Dim_i.Length \times Dim_i.Stride
\end{aligned}
\end{equation}
 \vspace{-3mm}

\subsection{Cache Architecture.}\label{subsec:cachearchitecture}

\begin{figure}[t]
    \centering
    \includegraphics[width=.43\textwidth]{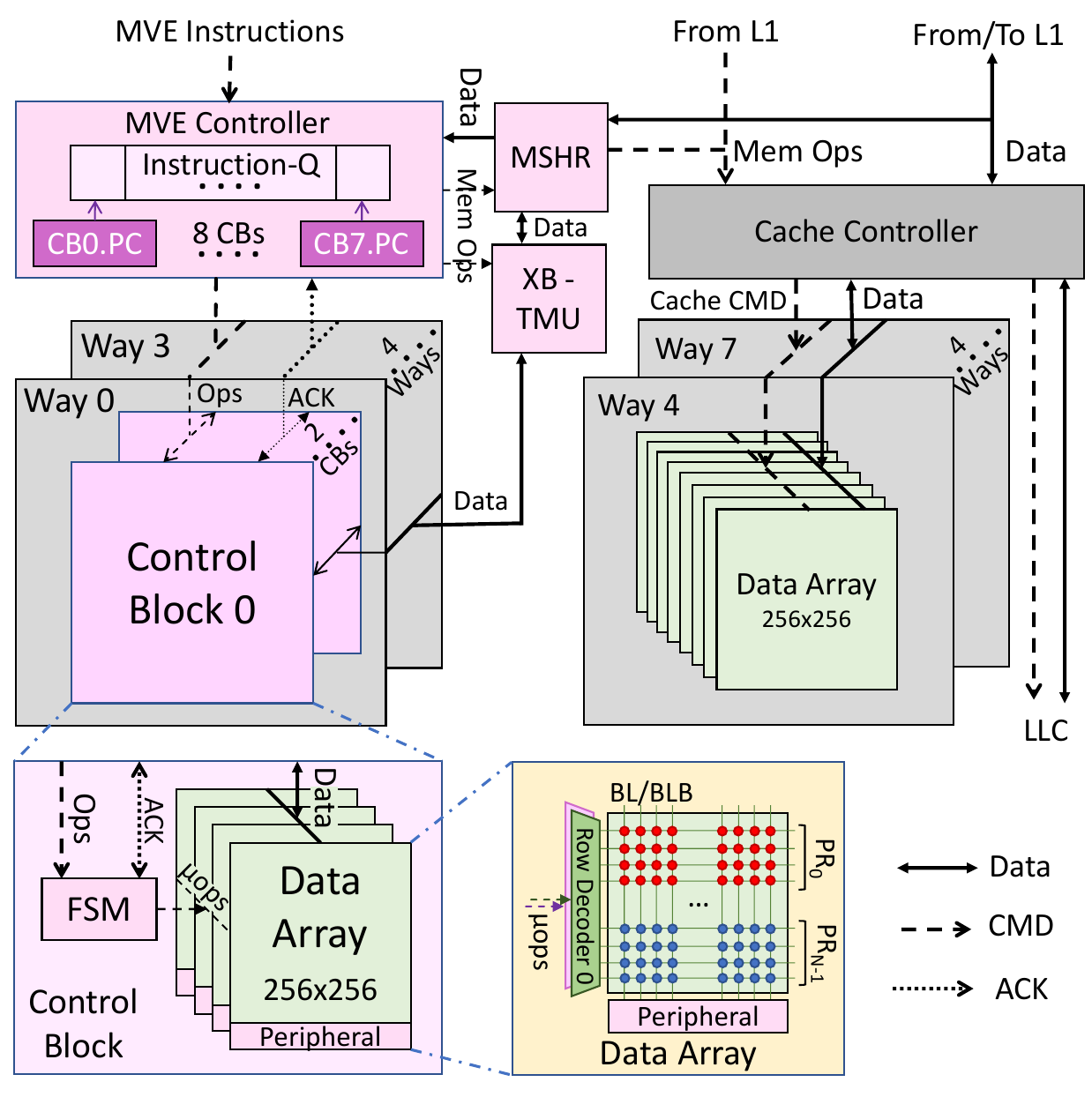}
    \vspace{-4mm}
    \caption{\gpic{} adds pink modules to the Cache Architecture.}
    \label{fig:cachearch}
\end{figure}

Figure~\ref{fig:cachearch} shows the new modifications to the cache to enable the execution of \gpic{} in the pink color:

\textbf{\gpic{} controller} receives \gpic{} instructions from the core in the program order and enqueues them to the \textit{Instruction-Q}.
When decoding a config instruction, \gpic{} controller sets the CR values similar to the Address Decoder in the core.
The logical registers are flattened out to the physical registers based on the dimension lengths, mapped to the \textit{Data Arrays}, and issued to the \textit{Finite State Machines} (FSM).
FSMs further decode the instructions to micro-operations ($\mu$ops) that control the \textit{Row Decoders} and \textit{Logic Peripherals} of SRAM arrays.

\textbf{Control blocks.}
Augmenting each SRAM array with an independent FSM results in high area overhead. 
Therefore, we use a coarser granularity of four SRAM arrays for a \textit{Control Block} (CB) and use a single FSM for each CB~\cite{dualitycache}.
Due to the \textit{dimension-level mask operations}, some CBs can be masked off of a \gpic{} instruction.
Therefore, \gpic{} controller orchestrates the non-blocking execution of CBs by keeping a \textit{mask bit-vector} that shows to which CBs each instruction must be issued.
If a CB is not masked off, \gpic{} controller issues the instruction to that CB. 

\gpic{} controller keeps a PC for each CB, which points to the instructions in the Instruction-Q.
When a CB completes an instruction, it interrupts the \gpic{} controller by raising the \textit{ACK} signal.
\gpic{} controller unsets the corresponding bit-vector of the completed instruction.
While increasing PC, \gpic{} controller jumps over the instructions whose bit-vector is masked off for that CB.
When the mask bit-vector of an instruction is all unset, it is dequeued from the Intrinsic-Q.
The \gpic{} ISA does not support any register-to-register permutation or shuffling, and any data transfer between SRAM arrays and bit-lines is through the memory (Table~\ref{tab:ISA}).
For that, we make sure there is only one load or store instruction executed in parallel in all CBs.
Thus, \gpic{} controller blocks on vector memory accesses until all CBs finish executing it.

\textbf{Strided and random memory access execution.}
While issuing a memory access to a CB, \gpic{} controller calculates the individual addresses using base address(es), stride modes, and CRs (Algorithm~\ref{alg:stridedmemoryaccess} and Equation~\ref{eq:randommemoryaccess}).
While there are opportunities to coalesce the memory accesses in the \gpic{} controller based on the stride semantics, we decided to simplify the controller and use Miss Status Holding Registers (MSHR) for this purpose.
The calculated addresses of each SIMD lane is sent to the MSHR, loaded from the regular half of the L2 cache, and delivered to the \textit{Transpose Memory Unit (TMU)}.
TMU is composed of 8T transpose bit-cells that can access data in both vertical and horizontal directions~\cite{neuralcache}.
TMU is sized to maintain a physical register for a CB (1024 elements).
The data words are routed to their correct vertical locations in the TMU SRAM cells using a \textit{crossbars (XB)}.
When all elements of a CB are received, the \gpic{} controller signals the corresponding CB to read the register bit slices from the TMU horizontally and write to the data arrays.
The store instructions require the opposite of the above process.

\vspace{-2mm}
\subsection{Discussion}

\textbf{Cache Coherency.}
Conventional cache coherency directories keep a single cache line state for the L1 and L2 caches of each core. 
However, in \gpic{}, the L2 cache accesses memory for in-cache loads and stores independently of the core and the L1 cache.
Consequently, cache levels are no longer coherent.
We observe that the baseline L2 cache is strictly inclusive and the tag array contains a \textit{presence} bit for each cache line, indicating if it is valid in the L1 cache~\cite{cortexa76}.
To maintain the coherency with the L1 cache, we enable the \gpic{} controller to check the \textit{presence} bit of the tag array.
Upon a cache hit for the in-cache memory access, the \gpic{} controller checks the \textit{presence} bit and evicts the cache line from the L1 cache if it is set.
An inclusive L2 cache miss guarantees that the data is not in the L1, ensuring coherency between the L1 and L2.

\textbf{What if in-cache computing is not needed?}
We carefully equip half of the cache ways~\cite{dualitycache} with in-cache computing while preserving their storage capabilities for scalar or SIMD applications.
\gpic{} does not permanently disable half of the cache banks to support the compute mode.
Cache banks operate as regular cache when not used for in-SRAM computing.
For workloads that do not benefit from long-vector execution, we default to normal CPU execution with regular cache usage.
To switch from the normal cache to in-cache computing for long-vector applications, \gpic{} flushes dirty cache lines.
Based on a prior heuristic of 50\% L2 cache lines being dirty~\cite{dirtylines}, this process takes less than 2\% of the execution time of the evaluated benchmarks.
On the other hand, switching compute-cache to normal-cache only requires modifying a CR with negligible slowdown to the workload.
\section{Evaluation Methodology}

\textbf{Benchmarks.}
We evaluate \gpic{} using 44 data parallel kernels of 12 real-world mobile libraries from the Swan benchmark suite~\cite{swan} as listed in Table~\ref{tab:benchmark}.
For a detailed analysis and comparison with prior work, we select 11 kernels with various dimensions and data-parallel patterns (Section~\ref{sec:patters}).

\begingroup
\setlength{\tabcolsep}{4pt} 
\renewcommand{\arraystretch}{1} 
\begin{table}[h]
\vspace{-4mm}
\scriptsize
  \renewcommand\arraystretch{1}
  \centering
  \caption{Evaluated Libraries}
    \vspace{-2mm}
  \scalebox{1}{
      \setlength{\tabcolsep}{2.5pt} 
      \renewcommand{\arraystretch}{1.4}
      \begin{tabular}{|c|c|c|c|c|}
    \hline
     \textbf{Domain} & \textbf{Library} & \textbf{\#Kernels} & \textbf{Dataset} & \textbf{Dim} \\
     \hline \hline
     Linear Algebra & Linpack~\cite{dongarra2003linpack} & 1 & 512K & 1D \\
     \hline
     Machine Learning & XNNPACK~\cite{xnnpack} & 2 & 156 CNN Layers & 2D \\
     \hline
     Signal Processing & CMSIS-DSP~\cite{cmsisdsp} & 3 & 192K & 1D \\
     \hline
     Video Processing & Kvazaar~\cite{hevc} & 4 & 1280$\times$720 & 3D \\
     \hline
     \multirow{3}{*}{\shortstack{Image \\ Processing}} & libjpeg~\cite{libjpeg} & 5 & \multirow{3}{*}{1280$\times$720} & \multirow{3}{*}{2-4D} \\
     \cline{2-3}
     & libpng~\cite{libpng} & 3 & & \\
     \cline{2-3}
     & libwebp~\cite{libwebp} & 7 & & \\
     \hline
     Graphics & Skia~\cite{skia} & 4 & 1280$\times$720 & 1-3D \\
     \hline
     Audio Processing & Webaudio~\cite{webaudio} & 5 & 32S$\times$44.1kHz & 1-3D \\
     \hline
     Data Compression & zlib~\cite{zlib} & 2 & 128KB & 1-2D \\
     \hline
     Cryptography & boringssl~\cite{boringssl} & 3 & 128KB & 1-2D \\
     \hline
     String/Network & Arm Optimized & \multirow{2}{*}{5} & \multirow{2}{*}{128KB} & \multirow{2}{*}{1-2D} \\
     Utilities & Routines~\cite{optroutines} & & & \\
     \hline
   \end{tabular}
   }
  \label{tab:benchmark}
\end{table}
\endgroup

\textbf{Mobile platform.} Table~\ref{tab:config} shows the baselines using Qualcomm Snapdragon 855 SoC: packed-SIMD (Arm Neon) execution of Arm Cortex-A76 core~\cite{cortexa76} and Qualcomm Adreno 640 GPU.
We use four L2 cache ways for data storage and four cache ways (32 SRAM arrays) for in-cache computing.

\begin{table}[h]
\vspace{-4mm}
  \scriptsize
  \centering
  \caption{Qualcomm Snapdragon 855 Baselines}
  \vspace{-2mm}
  \scalebox{0.95}{
  \begin{tabular}{|c|c|}
    \hline 
    \textbf{Configuration} & \textbf{Detail} \\
    \hline \hline
    Scalar core & 2.8GHz, 4-way out-of-order, 128 entry ROB \\
    \hline
    Vector engine & 2 128-bit Advance SIMD units + crypto and FP16 ext \\
    \hline
    L1-I cache & 64KB, 4-way, 4 cycle latency, 20 MSHRs  \\
    \hline
    L1-D cache & 64KB, 4-way, 4 cycle latency, 20 MSHRs  \\
    \hline
    L2 cache & 512KB, 8-way, Private, Inclusive, 12 cycle latency, 46 MSHRs \\
    \hline
    LLC & 2MB, 8-way, Shared, Inclusive, 31 cycle latency, 64 MSHRs/way  \\
    \hline
    \hline
    \gpic{} & 32 8-KB SRAM Arrays, 4-SA CB, 2KB Intrinsic-Q \\
    \hline
    \hline
    GPU & 2 cores, 384 ALUs, 685MHz, 1MB on-chip memory \\
    \hline
   \end{tabular}
   }
  \label{tab:config}
\end{table}

\textbf{Performance modeling.}
We developed a synthetic \gpic{} intrinsic library that emulates 29 intrinsics for 6 data types and validated the functional correctness by comparing the outputs with the Arm Neon kernels.
To generate instruction traces, we implemented a DynamoRIO client~\cite{dynamorio} that dumps the dynamic instructions on an Armv8.2-A server CPU (same as the mobile processor) while replacing function calls to the intrinsic library with their corresponding opcodes.
Our compilation infrastructure performs register allocation and instruction scheduling to the in-cache registers and \gpic{} instructions.
We then built a trace-driven, cycle-accurate simulator, faithfully modelling the micro-architecture of an Arm processor with the configurations of Table~\ref{tab:config} and cache architecture components explained in Section~\ref{sec:design} and Figure~\ref{fig:cachearch}.
We use the bit-serial in-SRAM instruction latency of Duality Cache~\cite{dualitycache}.
Our simulator injects memory accesses to the Ramulator~\cite{ramulator} to model the memory latency and bandwidth.

\ignore{
like instruction decode (config registers), dispatch and retire (register renaming), issue (data dependency), LSQ (memory dependency), core-cache latency and bandwidth, and cache
We ensure the correctness of the consistency model and coherency protocol using numerous assertions.
Our trace-driven simulator predicts the execution time of Arm Neon kernels on par with a \textit{full-system} cycle-accurate simulator~\cite{6322869}. 
}


\textbf{Power and area modeling.} We use the bit-serial in-SRAM computing energy parameters from \cite{neuralcache} and evaluate cache access energy using CACTI~\cite{cacti}.
\textit{Dynamic} CPU and GPU energy consumption are measured using \textit{Batterystats}~\cite{batterystats} and \textit{Trepn profilter}~\cite{trepn}.
We take the TMU, XB, FSM, and Peripheral area from prior work~\cite{neuralcache,dualitycache} and scale the numbers to MVE architecture requirements.
We implemented the controller and address decoder in RTL and synthesized it with a 15nm CMOS library.
MSHR area is evaluated using CACTI~\cite{cacti}.
Scalar CPU core and GPU area are estimated using die shots~\cite{dieshot1, dieshot2}, and Arm Neon area is generously evaluated using a vector engine design~\cite{aravec}.
We carefully scale all area and energy numbers to 7nm using equations of~\cite{techscale}.

\textbf{Prior work.} For Duality Cache~\cite{dualitycache}, we implement kernels in CUDA and run them on GPU Ocelot emulator \cite{gpuocelot} to capture dynamic PTX trace. We calculate compute cycles using our in-cache operation latencies and simulate memory accesses by Ramulator~\cite{ramulator}.
For BH, BP, and AC schemes, we implement kernels with various algorithms optimized for their number of SIMD lanes and in-cache registers.
We ignore different operational frequencies and architectural contributions of prior work to demonstrate the benefits of MVE for various in-SRAM computing schemes in a similar hardware configuration.
The compiler and trace-driven simulator are configured with in-cache register count, number of SIMD lanes per SRAM array, and compute and data movement latencies based on prior works~\cite{bsbppincache,eve,cape}.
To compare \gpic{} with RISC-V RVV~\cite{riscv}, we implement workloads using optimized algorithms for only 1D vector instructions.




\section{Results}
\label{sec:results}

\vspace{-2mm}

\subsection{Performance and Energy Analysis}~\label{subsec:perf_energy}

\vspace{-4mm}

\begin{figure}[t]
    \centering
   \includegraphics[width=0.49\textwidth, trim={0.1cm 0.2cm 0.1cm 0.1cm}, clip]{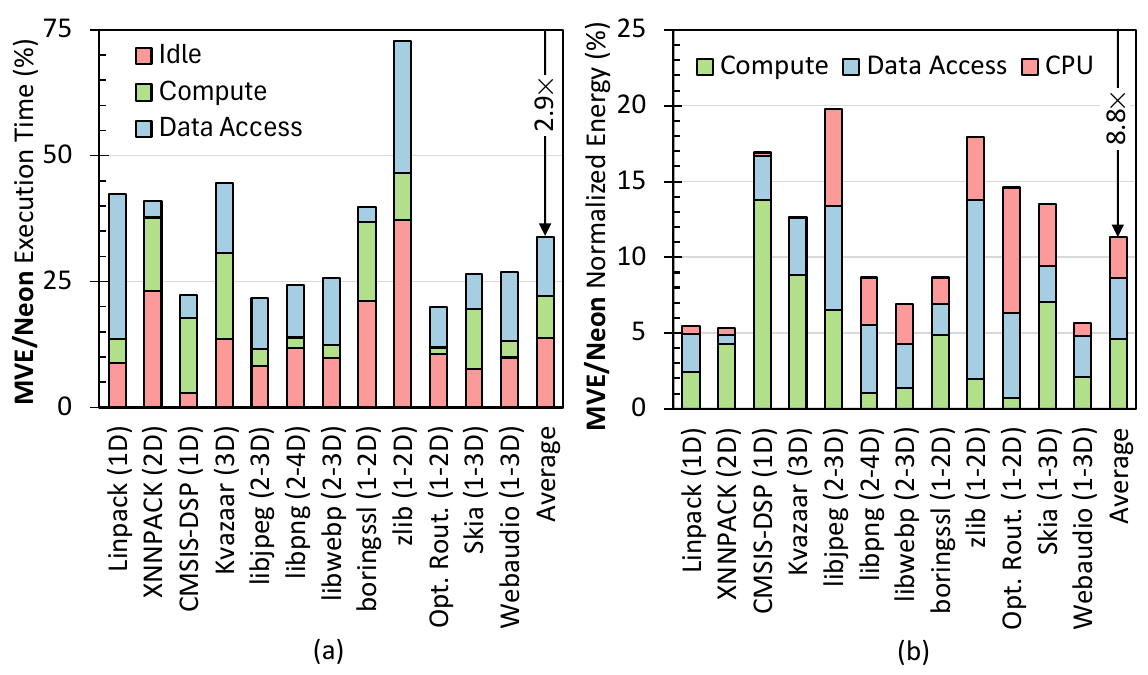}
   \vspace{-6mm}
   \caption{(a) Execution time and (b) energy consumption of \gpic{} normalized to packed-SIMD Neon model.}
    \label{fig:performance_vs_cpus}
\end{figure}

\textbf{\gpic{} vs. Arm Neon.}
Figure~\ref{fig:performance_vs_cpus}(a) shows the execution time of \gpic{} normalized to Arm Neon.
Due to the limited space, we only show the average for all kernels of a library.
\gpic{} outperforms Neon by 2.9$\times$ across all evaluated mobile kernels, showing that \gpic{} efficiently exploits the higher compute throughput of in-cache computing.
The execution time is classified as 40.4\% idle, 24.7\% compute, and 34.8\% data access time, on average.
\gpic{} performance improvement increases in kernels with lower precision (see Section~\ref{subsec:sensitivity}).
So, \gpic{} significantly outperforms Neon for \textit{libpng}, \textit{libjpeg}, \textit{libwebp} (image codecs), \textit{Skia} (graphics), and \textit{Arm optimized routines} (string/network utility) libraries, which operate on 8-bit pixel values and characters.

The idle time is when the control blocks (CBs) have no \gpic{} instructions to execute.
It increases when the portion of scalar instructions is higher.
For example, the \textit{Adler32} kernel of the \textit{zlib} library (compression) requires the reduction kernel (explained in Section~\ref{sec:patters}).
In the vertical reduction based on the tree algorithm, many CBs are idle.
The scalar core reduces the elements from 256 to 1, while the CBs are idle.
The idle time drops the \textit{zlib} performance improvement to only 37$\%$.

\gpic{} improves the energy consumption by 8.8$\times$ on average as shown in Figure~\ref{fig:performance_vs_cpus}(b).
Significant energy reduction in \gpic{} is because of two reasons:
(a) the long-vector execution model effectively packs numerous operations into one multi-dimensional instruction, reducing CPU instructions and energy.
(b) the in-SRAM computing operations are energy-efficient~\cite{computecaches}.
In addition, in-cache computing prevents the data transfer from the L2 cache to the core's register files and back.

\textbf{\gpic{} vs. GPU and DSP.}~\label{subsec:gpu_and_dsp}
Mobile Application Processors integrate specialized accelerators like GPU, DSP (NPU), ISP, and SPU to accelerate \textit{specific application domains} such as multimedia processing, machine learning, image signal processing, and security.
In contrast, \gpic{} is a \textit{general-purpose} approach to flexibly accelerate all application domains.

\begin{figure}[t]
    \centering
    \includegraphics[width=0.49\textwidth, trim={0.1cm 0.5cm 0.1cm 0.3cm}, clip]{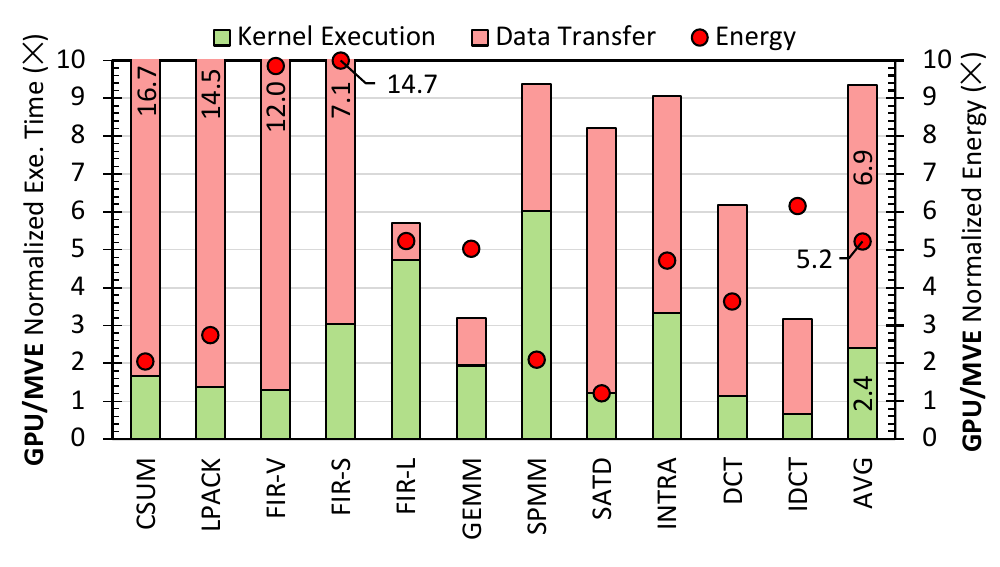}
    \vspace{-7mm}
    \caption{Performance (primary Y-axis) and Energy (secondary Y-axis) consumption of Adreno 640 GPU normalized to \gpic{}.} 
    \label{figure:gpu_comparison}
\end{figure}

Figure~\ref{figure:gpu_comparison} compares the execution time and energy consumption of \gpic{} with Adreno 640 GPU.
Despite 13.6$\times$ lower 32-bit integer MAC throughput, \gpic{} outperforms GPU by 9.3$\times$ and reduces energy by 5.2$\times$ because of two primary reasons:

(a) \textit{Data transfer} incurs a significant overhead for GPUs.
To reduce this cost, we utilize unified physical memory following the Snapdragon OpenCL Optimization Guide~\cite{openclopt}.
While we do not account for the cost of memory allocation, transferring data from complex C++ objects to pinned C array pointers in the unified memory region incurs a substantial overhead, taking 6.9$\times$ longer than the average execution time of \gpic{}.

(b) After discounting the data transfer time, \gpic{} still provides an average speedup of 2.4$\times$.
This is attributed to the high \textit{kernel launch overhead} of OpenCL Runtime library~\cite{swan} and core-GPU communication via the system fabric.
Similarly, DSP experiences kernel launch overhead, including latencies introduced by \textit{ADSPRPC}, \textit{skel}, and \textit{kernel object} transfers, \textit{stub/skel} parameter de/serialization, and \textit{FastRPC} remote calls through the system fabric~\cite{hexagon,hexagonsdk}.
Conversely, \gpic{} avoids these overheads by leveraging the fine-grained scalar-vector instruction interleaving of the SIMD execution model.

Finally, note that we compare a single-core \gpic{} performance with the GPU.
Augmenting all 8 cores of Snapdragon 855 SoC with in-cache computing would further increase the performance superiority of \gpic{} compared to a single GPU.

\begin{figure}[t]
    \centering
    \vspace{-3mm}
    \includegraphics[width=0.49\textwidth, trim={0.1cm 0.2cm 0.1cm 0.1cm}, clip]{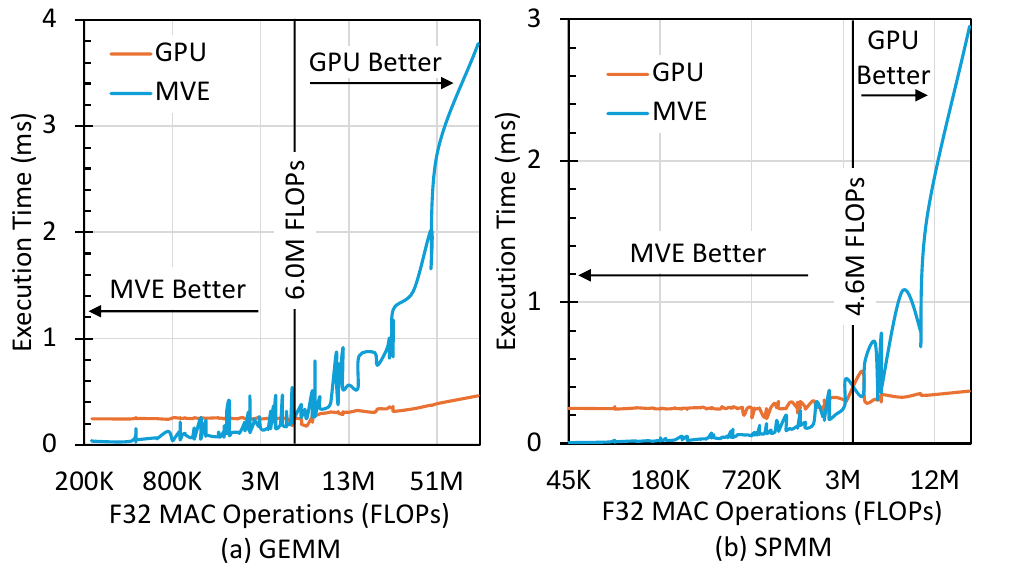}
    \vspace{-7mm}
    \caption{GEMM and SpMM execution time of \gpic{} and Adreno 640 GPU for different operation counts (matrix sizes).} 
    \label{figure:gemm_spmm}
\end{figure}

To compare \gpic{} and GPU in more depth, we take the OpenCL \textit{GEMM} and \textit{SpMM} implementations from CLBlast~\cite{clblast} and clSPARSE~\cite{clsparse} libraries, and evaluate them for different matrix sizes employed in 152 layers of 13 mobile Convolutional Neural Networks~\cite{xnnpack} using Figure~\ref{figure:gemm_spmm}.
\gpic{} outperforms GPU up to matrix multiplications with 6.0M and 4.6M FLOPs.
In large problem sizes, the higher throughput of GPU offsets its kernel launch overhead.
On the other hand, \gpic{} requires only 8K DLP to fully utilize the SIMD lanes, so it is preferred for smaller problem sizes.

\subsection{Multi-dimensional ISA Benefits}~\label{subsec:isa_micro_perf}
\vspace{-4mm}

\begin{figure}[t]
    \centering
    \includegraphics[width=0.49\textwidth]{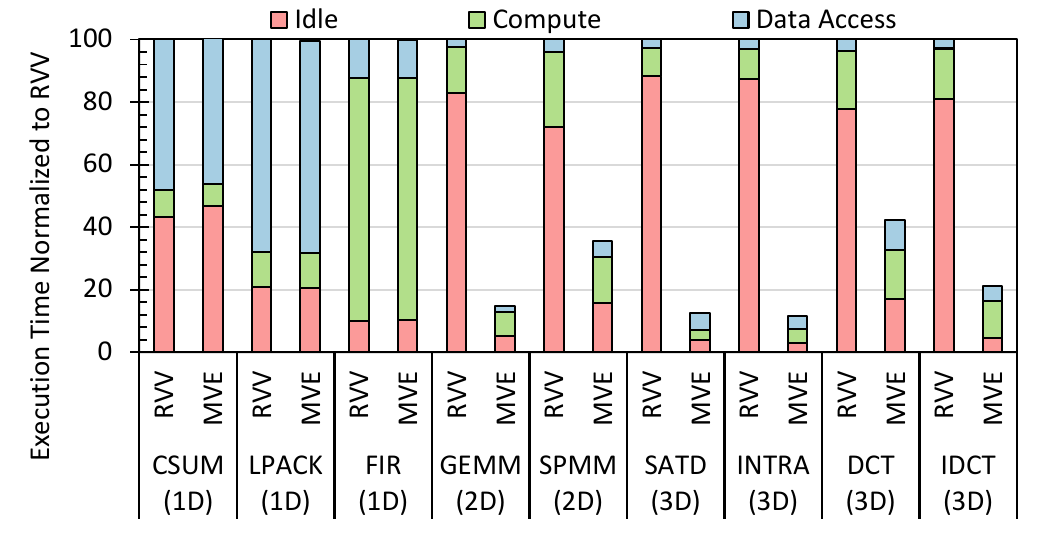}
    \vspace{-6mm}
    \caption{\gpic{} Performance compared to RISC-V RVV~\cite{riscv} when targeting the same underlying bit-serial in-cache computing engine with 8K SIMD lanes.}
    \vspace{-4mm}
    \label{fig:MVE_vs_RVV}
\end{figure}

\begin{figure}[t]
    \centering
    \includegraphics[width=0.49\textwidth]{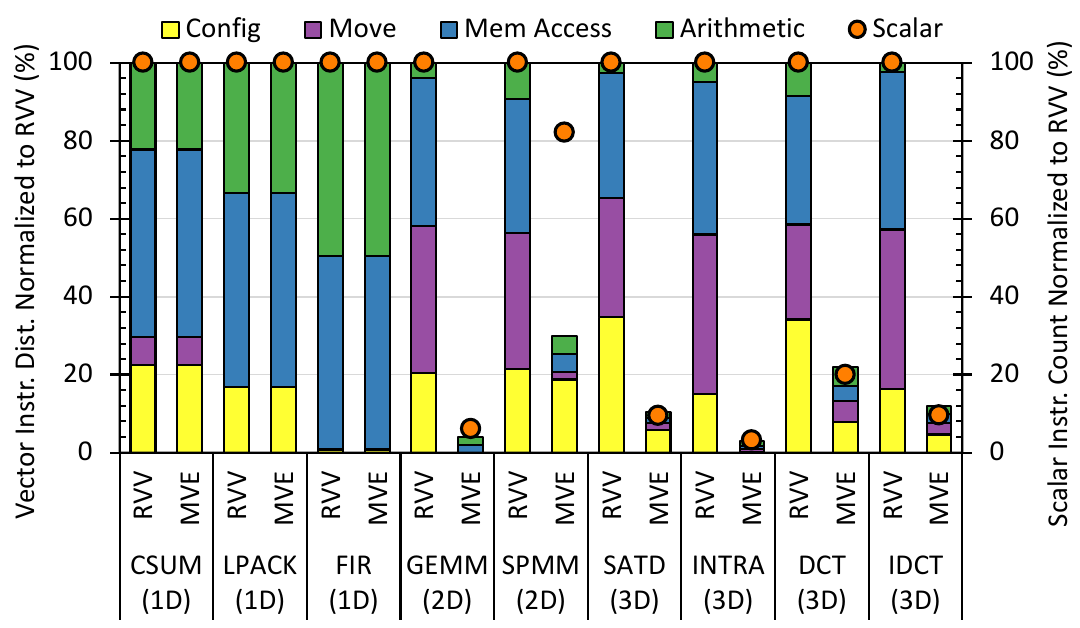}
    \vspace{-8mm}
    \caption{\gpic{} and RISC-V RVV~\cite{riscv} dynamic vector instruction distribution (left Y-axis) and dynamic scalar instruction count (right Y-axis).}
    \label{fig:MVE_vs_RVV_instr}
\end{figure}

Figure~\ref{fig:MVE_vs_RVV} and Figure~\ref{fig:MVE_vs_RVV_instr} compare the performance and dynamic instruction count of \gpic{} with a 1D long-vector ISA extension such as RISC-V RVV~\cite{riscv} while using the same underlying bit-serial in-cache computing engine.
We select various kernels for this comparison with different dimensions from 1D to 3D.
\gpic{} significantly reduces the idle time of the kernels that require multi-dimensional accesses, and improves the performance by 2.0$\times$ on average across all selected kernels.
To mimic a multi-dimensional access using RVV, we use predicate operations to mask 1D strided memory accesses and (un)pack each 1D segment of data structure separately in a piece of vector registers.
This approach generates $\frac{\#SIMD\ Lanes}{length(1D\ segment)}$ config (mask), partial 1D memory accesses, and move instructions for each multi-dimensional memory access.
Figure~\ref{fig:MVE_vs_RVV_instr} (left Y-axis) illustrates that \gpic{} reduces the number of dynamic vector instructions by 2.3$\times$, on average.

\begin{figure*}[t]
    \centering
    \includegraphics[width=0.99\textwidth]{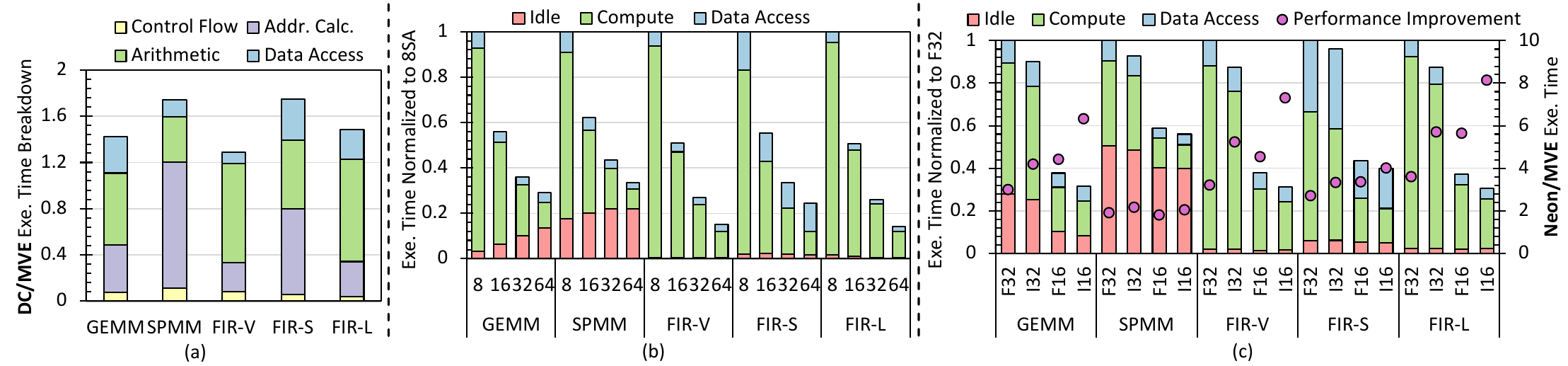}
    \vspace{-2mm}
    \caption{(a) Performance comparison with Duality Cache~\cite{dualitycache}. (b) Performance scalability using various numbers of SRAM Arrays (SA), and (c) Performance sensitivity to the data precision.}
    \vspace{-7mm}
    \label{fig:DC_SCALE_SENSE}
\end{figure*}

Across various GEMM kernels in the XNNPACK library~\cite{xnnpack}, RVV significantly increases dynamic vector instructions in smaller output matrices (less 1D parallelism) compared to \gpic{}.
For instance, a \texttt{128$\times$3136} output matrix from MobileNet V1~\cite{mobilenet} requires approximately $\frac{8192}{3136} \approx 3$ mask, partial memory accesses, and move instructions across three 1D segments (output rows) to occupy 8K SIMD lanes.
This results in a 5.3$\times$ increase in the vector instruction count compared to \gpic{}.
Another \texttt{122$\times$784} output matrix from ShuffleNet V2~\cite{shufflenetv2} involves $\frac{8192}{784} \approx 11$ 1D segments, leading to a 13.0$\times$ increase in the vector instructions.


Additionally, more partial memory accesses require more scalar address calculation instructions.
Hence, Figure~\ref{fig:MVE_vs_RVV_instr} (right Y-axis) shows that \gpic{} reduces the number of dynamic scalar instructions by an average of 2.0$\times$ compared to RVV.


\subsection{Comparing In-SRAM Technologies}

\begin{figure}[t]
    \centering
    \includegraphics[width=0.45\textwidth]{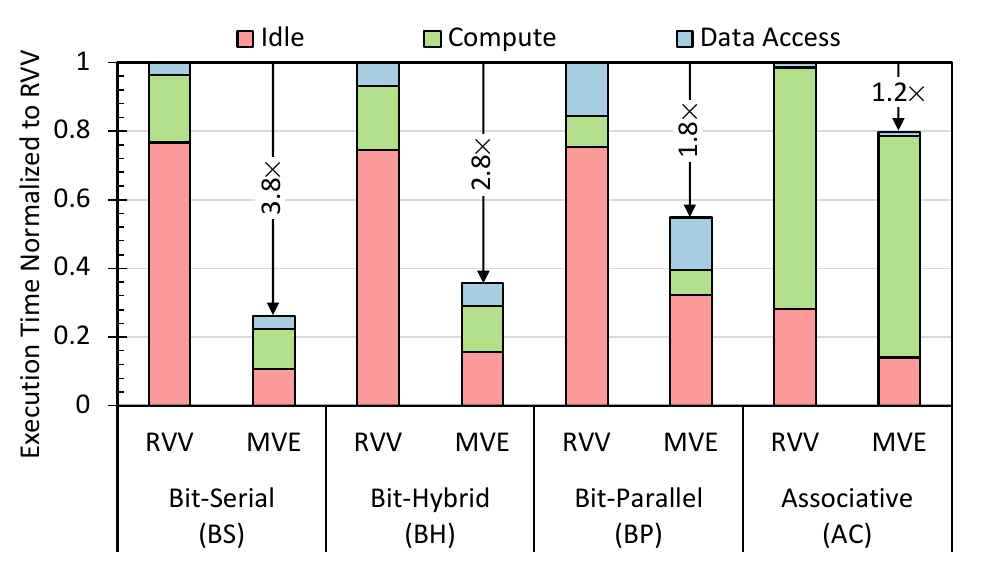}
    \caption{Performance improvement of \gpic{} compared to RISC-V RVV~\cite{riscv} for different in-SRAM computing schemes employed in VRAM~\cite{bsbppincache}, EVE~\cite{eve}, and CAPE~\cite{cape}.}
    \label{fig:MVE_vs_RVV_ALLIS}
\end{figure}

\textbf{VRAM~\cite{bsbppincache}, EVE~\cite{eve}, and CAPE~\cite{cape}}
employ the RVV ISA extension to instruct a bit-parallel (BP), bit-hybrid (BH), and Associative Computing (AC) in-SRAM engine.
Figure~\ref{fig:MVE_vs_RVV_ALLIS} shows that \gpic{} improves the performance of all in-SRAM computing schemes when compared with RVV.
Numerous partial memory accesses of RVV drop CB utilization of the BS to 23\%, BH to 25\%, and BP to 24\%.
\gpic{} efficiently encode multi-dimensional memory accesses in only one vector instruction, increasing the CB utilization of BS to 60\%, BH to 55\%, and BP to 41\%.
Among these schemes, BP still shows relatively lower CB utilization. 
This is because BP enjoys less compute latency at the cost of fewer SIMD lanes.
Therefore, the in-cache engine is faster in consuming the long-vector instruction, which increases the pressure on the core to issue instructions at a higher rate.
Less idle time improves the performance of BS, BH, and BP by 3.8$\times$, 2.8$\times$ and 1.8$\times$.
While \gpic{} also cuts down the idle time of AC by 2.0$\times$, this in-SRAM computing scheme does not benefit significantly from a multi-dimensional ISA.
This is because the latency of the AC arithmetic operations is 4-8$\times$ higher than BS, hence, 70\% of AC execution time is spent on arithmetic operations.

\textbf{Duality Cache~\cite{dualitycache}.}
Figure \ref{fig:DC_SCALE_SENSE} (a) demonstrates that \gpic{}, using a SIMD model, achieves an average performance improvement of 1.5$\times$ over the prior SIMT-based in-cache computing design, Duality Cache~\cite{dualitycache}.
The performance improvement is primarily due to two reasons:
\textit{First,} \gpic{} reduces compute cycles as it requires fewer in-SRAM computing operations.
In the SIMT model, all operations (control flow, address calculation, and arithmetic) are offloaded to the in-SRAM computing engine and performed individually by all SIMT lanes.
In contrast, \gpic{} executes control flow and base address calculations once on the scalar core, while the cache controller computes memory addresses for each SIMD lane using the stride semantics of \gpic{} memory accesses.
Therefore, the SIMT model removes the idle cycles of the in-SRAM computing engine at the cost of more in-SRAM operations.
So, the SIMT model is ideal for large server-class in-SRAM engines, where control-flow divergence poses a challenge, while the SIMD model is better suited for small mobile caches that prioritize low latency.
\textit{Second,} data access time of the SIMT model takes 1.6$\times$ longer than \gpic{}.
This is because the SIMT model keeps all scalar and vector variables in the scarce in-cache physical registers, leading to frequent register spills and fills.
These operations are particularly costly due to the large number of data elements in the in-cache physical registers (8K elements).

\subsection{Scalability Analysis}~\label{subsec:scalability}
Figure~\ref{fig:DC_SCALE_SENSE} (b) shows the performance scalability of \gpic{} with different SRAM array counts.
These baselines are designed by keeping the 512 KB L2 cache intact and augmenting the core with an additional 8 to 64 SRAM arrays dedicated for \gpic{}.
This result shows that increasing SRAM arrays by 8$\times$ increases \gpic{} performance from 3.0$\times$ (SpMM) to 6.7$\times$ (FIR-L).
Idle time is the primary bottleneck for scaling the performance of \gpic{} for the massive server-class caches.

\subsection{Sensitivity to Bit Precision}\label{subsec:sensitivity}
Figure~\ref{fig:DC_SCALE_SENSE} (c) shows the impact of bit precision on the execution time of \gpic{} (left Y-axis) and its performance improvement over Neon (right Y-axis) with two takeaways:
\textit{First,} \gpic{} shows greater performance improvement over Neon at lower precisions, indicating that in-SRAM computing benefits more from lower bit precisions than CMOS-based processors.
This is because in-SRAM computing relies on a bit-serial paradigm with an arithmetic complexity of $\mathcal{O}(precision^2)$.
So, bit precision has a quadratic effect on the throughput.
In contrast, Neon ASIMD units achieve linear scaling with lower bit precision, as they can pack more data elements into a SIMD register.
\textit{Second,} idle cycles in \gpic{} do not decrease significantly with lower precision.
This is because CBs execute lower-precision in-SRAM instructions much faster, outpacing the rate of issuing instructions by the core.

\subsection{Area Analysis}~\label{subsec:overhead_analysis}
Table~\ref{tab:area} evaluates the area overhead of \gpic{} compared to Snapdragon 855 mobile SoC.
While augmenting a scalar data path with the Neon register files and ALUs incur 16.3\% area overhead, \gpic{} transforms the cache structures to a long-vector engine with a negligible 3.6\% area overhead \textit{to the scalar core}.
This area overhead is similar to the findings of the prior work, \textit{e.g.}, 3.5$\%$~\cite{dualitycache} and 2$\%$~\cite{neuralcache}.
\gpic{} area is dominated by FSM controllers, which decode \gpic{} instructions to the $\mu$ops.
This table further shows the significantly larger area of GPU.
Hence, a general-purpose \gpic{} design offers both performance and energy benefits over dedicated hardware components at a negligible silicon area, making it an attractive solution for the cost-sensitive mobile industry.

\begin{table}[h]
 \vspace{-4mm}
  \centering
  \caption{Overhead to the Scalar Core Area (1.07$mm^2$~\cite{dieshot1})}
 \vspace{-3mm}
  \begin{tabular}{|c|c|c|c|c|}
    \hline
    \multicolumn{2}{|c|}{\textbf{Module}} & \textbf{Source} & \textbf{Area ($mm^2$)} & \textbf{Overhead (\%)} \\
    \hline
    \hline
    \multicolumn{2}{|c|}{Arm Neon} & \cite{aravec} & 0.1741 & \textbf{16.321} \\
    \hline
    \hline
    \multirow{8}{*}{\vrotate{\textbf{\gpic{}}}} & Controller & RTL & 0.0043 & 0.409 \\
    \cline{2-5}
    & MSHR & CACTI & 0.0018 & 0.168 \\
    \cline{2-5}
    & TMU & ~\cite{neuralcache} & 0.0053 & 0.498 \\
    \cline{2-5}
    & XB & ~\cite{dualitycache} & 0.0039 & 0.367 \\
    \cline{2-5}
    & FSM & ~\cite{dualitycache} & 0.0123 & 1.154 \\
    \cline{2-5}
    & Peripheral & ~\cite{dualitycache} & 0.0063 & 0.590 \\
    \cline{2-5}
    & Address Decoder & RTL & 0.0042 & 0.399 \\
    \cline{2-5}
    & Total & -- & 0.0382 & \textbf{3.588} \\
    \hline
    \hline
    \multicolumn{2}{|c|}{Adreno 640 GPU} & ~\cite{dieshot2} & 11.1908 & \textbf{--} \\
    \hline
   \end{tabular}
  \label{tab:area}
  \vspace{-2mm}
\end{table}

\section{Related Work}

\textbf{Domain-specific in-cache computing.} In-cache computing is specifically studied for different domains such as Machine Learning \cite{computecaches, neuralcache, bitprudent}, Automata Processing \cite{cacheautomation, eap, impala}, Image Processing \cite{imageprocessing}, and Data Encryption \cite{sealer}. However, domain-specific acceleration requires extra effort for customizing data layout of the cache and dataflow between core and in-cache elements. A general-purpose execution model such as \gpic{} lifts this burden from the programmers, unleashing the power of in-cache computing for other domains.

\textbf{General-purpose ISA design for in-cache computing.}
Several works have extended the scope of in-cache computing for general applications. We already comprehensively discuss Duality Cache~\cite{dualitycache}, CAPE~\cite{cape} and EVE~\cite{eve}.
Kooli \textit{et al.}~\cite{instructioncode, kooli22} propose a general-purpose SIMD model for in-cache computing, integrating the execution flow with the CPU.
They also introduce an in-place shuffle operation to accelerate cryptography, image processing, \textit{etc.}
Blade~\cite{blade} proposes circuit-level solutions for in-cache computing to operate at high-frequency and low-voltage of mobile processors.
The architectural contributions of these prior works are orthogonal to this paper.
\gpic{} could be employed to enhance the performance of all prior In-SRAM architectures (Section~\ref{subsec:isa_micro_perf}).

Prior work demonstrates the multi-dimensional access pattern of in-memory databases where the data can be accessed column and row-wise in a logical linear address space.
GSDRAM~\cite{gsdram} proposes changes in the DRAM controller to fetch power-of-two strided access patterns independently from multiple DRAM chips, improving DRAM bandwidth utilization.
MDACache~\cite{mdacache}, RC-NVM~\cite{rcnvm}, and MVC~\cite{mvc} modify the computing stack and cache architecture to enable vertical and horizontal access to the non-volatile memories, improving cache line utilization.
These proposals can complement \gpic{} and be employed to enhance the DRAM memory bandwidth.

\textbf{Mobile vector processing.}
Big.VLITTLE~\cite{bigvlittle} re-purposes little cores in the big.LITTLE architecture of mobile processors as a decoupled vector processing engine without adding area-expensive components such as vector register files and wide execution pipelines.
\gpic{} supports a larger SIMD capacity tightly integrated into the scalar core by repurposing half of the processor's L2 cache.
PUMICE~\cite{pumice} proposes out-of-order dispatch of in-cache long-vector instructions to hide the command distribution overhead of CAPE~\cite{cape}.
\gpic{} architecture can also employ this technique to reduce idle time.




\textbf{Long-vector and Matrix processing.}
The emergence of multimedia extensions two decades ago inspired designs to enhance the DLP exposed to SIMD engines.
MOM~\cite{mom,threedim} proposed 2D SIMD instructions while supporting flexible vector length and stride in the second dimension.
Breeze~\cite{breeze} and CSI\cite{csi} proposed CISC-based instructions to encode all operations, data access patterns, and nested loop semantics for offloading to a streaming media processor.
Motorola 56000~\cite{motorola_dsp56000} and Texas Instruments TMS320C5x~\cite{ti_tms320c5x} DSPs employ sophisticated DO loop and strided address calculation instructions to reduce the control flow and address calculation overhead.
More recently, Nvidia server GPUs have incorporated Tensor Memory Accelerator (TMA) units~\cite{tma} capable of bulk copying 1-5D tensors from global memory to on-chip shared memory.
Further, ISA extensions are proposed for machine learning workloads in server processors. Intel Advanced Matrix Extension (AMX)~\cite{intelamx} and Arm Scalable Matrix Extension (SME)~\cite{armsme} are 2D matrix ISA extensions, enabling high-performance 1K-bit and 2K-bit matrix operations. AMX has two main components: specialized loads to fixed-size tiles composed of eight 2D registers, each 1KB in size, and Tile Matrix Multiplication Units, which perform matrix-multiply computations.

As opposed to these designs, \gpic{} is a \textit{long-vector ISA} extension specifically targeting tightly coupled in-cache computing engines for mobile cores and capturing the semantics of data-parallel computation across multiple dimensions systematically. Besides, \gpic{} encodes irregular computations using random memory accesses and efficient dimension-level masked execution tailored towards mobile workloads.



\section{Conclusion}

This work introduces a long-vector \underline{M}ulti-dimensional \underline{V}ector ISA \underline{E}xtension (\gpic{}) that abstracts away the in-cache computing as a tightly-integrated long-vector processing engine.
\gpic{} enables the high utilization of numerous SIMD lanes using multi-dimensional strided and random memory accesses, and efficient dimension-based masked execution.
Using this ISA extension, \gpic{} repurposes half of the L2 mobile caches for in-cache computing with a $3.6\%$ area overhead.
The long-vector in-cache architecture offers $2.9\times$ performance improvement and $8.8\times$ energy reduction, compared to a commercial mobile CPU vector baseline.
Compared to the RISC-V RVV, \gpic{} improves the in-cache engine utilization from 23\% to 60\% and improves the performance by $3.8\times$ when targeting the same underlying bit-serial in-cache computing engine.

\section*{Acknowledgements}
\vspace{-1mm}
This work was supported in part by the NSF under the CAREER-1652294 and NSF-1908601 awards and by JST PRESTO Grant Number JPMJPR22P7.

\appendix
\section{Artifact Appendix}

\subsection{Abstract}

The artifact of this paper includes the in-cache computing simulator for \gpic{} ISA, various implementations of the benchmarks, and instrumentation tools for baseline evaluation.

\subsection{Artifact check-list (meta-information)}

{\small
\begin{itemize}
  \item {\bf Program:} Scalar, Arm Neon, OpenCL, CUDA, \gpic{}, and RVV implementations of evaluated Swan~\cite{swan} benchmarks (Table~\ref{tab:benchmark_impl}).
  \item {\bf Compilation:} Android NDK r23c for Arm Neon, Adreno OpenCL SDK v1.5 for mobile GPU, Clang 3.4+ for simulator, and CMake 3.7+ for DynamoRIO.  
  \item {\bf Software:} Git LFS.
  \item {\bf Hardware:} Samsung Galaxy S10e for the packed-SIMD Neon and Adreno GPU baselines, an Armv8.2-A machine for generating the simulation traces, and any system for simulation.
  \item {\bf Output:} CSV files generated by \href{https://github.com/arkhadem/MVE/tree/main/benchmarks/scripts}{\blue{automation scripts}}.
  \item {\bf Experiments:} Arm Neon (mobile SIMD) and Adreno (mobile GPU) measurements on the mobile device, \gpic{} and RVV simulations for bit-serial (BS), bit-hybrid (BH), bit-parallel (BP), and associative computing (AC) in-cache computing schemes.
  \item {\bf Required disk space:} Approximately 65 GB
  \item {\bf Preparation time:} Approximately 1 hour
  \item {\bf Measurement/simulation time:} Approximately 24 hours
  \item {\bf Publicly available:} Yes, available on \href{https://github.com/arkhadem/MVE/tree/main}{\blue{GitHub}} and \href{https://zenodo.org/records/14352812}{\blue{Zenodo}}.
  \item {\bf Code license:} Yes, \href{https://github.com/arkhadem/MVE/blob/main/LICENSE}{\blue{MIT License}}.
  \item {\bf Workflow automation:} Yes, located in \href{https://github.com/arkhadem/MVE/tree/main/benchmarks/scripts}{\blue{benchmarks/scripts/}}.
\end{itemize}
}

\subsection{Overview}~\label{subsec:artifact_overview}
This artifact consists of two parts for reproducing the results presented in the following figures: 
(i) Figure~\ref{fig:performance_vs_cpus}(a) (\gpic{} vs. Arm Neon), 
(ii) Figures~\ref{figure:gpu_comparison} and~\ref{figure:gemm_spmm} (\gpic{} vs. Adreno GPU), and 
(iii) Figures~\ref{fig:MVE_vs_RVV},~\ref{fig:MVE_vs_RVV_instr}, and~\ref{fig:MVE_vs_RVV_ALLIS} (\gpic{} vs. RVV). 
The first part involves conducting measurements on a mobile device, while the second simulates the \gpic{} and RVV kernels for the BS, BH, BP, and AC in-cache computing schemes.
Both parts require access to specific hardware dependencies, particularly Arm-based machines. 
If you do not have access to these machines, we provide precomputed result files to bypass them.

Table~\ref{tab:benchmark_impl} shows all implementations \href{https://github.com/arkhadem/MVE/tree/main/benchmarks/src/libraries/}{\blue{available here}}. 
We provide both Scalar and Neon implementations for all libraries; however, only the Neon implementations are used for comparison with the vector units (Arm Neon). 
The \gpic{} implementation is also provided for all libraries.
We selected 10 kernels with varying dimensions and implemented the following versions:
(a) Using one-dimensional instructions only, for comparison with RVV,
(b) Using OpenCL, to evaluate performance on the Adreno mobile GPU, and
(c) Using CUDA, to study the performance of the Duality Cache (DC)~\cite{dualitycache}.
Note that this artifact does not include the reproduction of Figure~\ref{fig:DC_SCALE_SENSE} (a) (\gpic{} vs. DC), due to the complexities with DC's simulation infrastructure (GPU Ocelot~\cite{gpuocelot}). 

\begin{table}[h]
  \renewcommand\arraystretch{1.0}
  \centering
  \caption{Benchmark Implementations.}
  \begin{tabular}{|c|c|c|c|c|c|c|c|}
    \hline
    \multirow{4}{*}{\textbf{Library}} &
    \multirow{4}{*}{\textbf{\#Kernels}} &
    \multirow{4}{*}{\vrotate{\textbf{Scalar}}} &
    \multirow{4}{*}{\vrotate{\textbf{Neon}}} &
    \multirow{4}{*}{\vrotate{\textbf{\gpic{}}}} &
    \multirow{4}{*}{\vrotate{\textbf{RVV}}} &
    \multirow{4}{*}{\vrotate{\textbf{OpenCL}}} &
    \multirow{4}{*}{\vrotate{\textbf{CUDA}}} \\
    &&&&&&& \\
    &&&&&&& \\
    &&&&&&& \\
    \hline \hline
    Linpack & 1 & \cmark & \cmark & \cmark & \cmark & \cmark & \cmark \\
    \hline
    XNNPACK & 2 & \cmark & \cmark & \cmark & \cmark & \cmark & \cmark \\
    \hline
    CMSIS-DSP & 3 & \cmark & \cmark & \cmark & \cmark & \cmark & \cmark \\
    \hline
    Kvazaar & 4 & \cmark & \cmark & \cmark & \cmark & \cmark & \cmark \\
    \hline
    Arm Opt. & 1 & \cmark & \cmark & \cmark & \cmark & \cmark & \cmark \\
    \cline{2-8}
    Routines & 4 & \cmark & \cmark & \cmark & \xmark & \xmark & \xmark \\
    \hline
    libjpeg & 5 & \cmark & \cmark & \cmark & \xmark & \xmark & \xmark \\
    \hline
    libpng & 3 & \cmark & \cmark & \cmark & \xmark & \xmark & \xmark \\
    \hline
    libwebp & 7 & \cmark & \cmark & \cmark & \xmark & \xmark & \xmark \\
    \hline
    Skia & 4 & \cmark & \cmark & \cmark & \xmark & \xmark & \xmark \\
    \hline
    Webaudio & 5 & \cmark & \cmark & \cmark & \xmark & \xmark & \xmark \\
    \hline
    zlib & 2 & \cmark & \cmark & \cmark & \xmark & \xmark & \xmark \\
    \hline
    boringssl & 3 & \cmark & \cmark & \cmark & \xmark & \xmark & \xmark \\
    \hline
   \end{tabular}
  \label{tab:benchmark_impl}
\end{table}


\textbf{How to access artifact?}
All the necessary codes, tools, and scripts are available in our \href{https://github.com/arkhadem/MVE/tree/main}{\blue{GitHub repository}}. This repository also includes large dynamic instruction and data flow graph files (approximately 2.5GB), which can be used to skip Steps 1 and 2 of Part 2 (Section~\ref{subsec:artifact_part2}). If you wish to clone the repository while downloading these large files, install \href{https://git-lfs.com/}{\blue{Git LFS}} before cloning. 
Otherwise, set the environment variable \texttt{GIT\_LFS\_SKIP\_SMUDGE=1} before cloning.

\subsection{Part 1 -- Mobile Measurements}

\textbf{Hardware dependencies.} This step requires a Samsung Galaxy S10e device equipped with the Snapdragon 855 SoC. 
If you do not have access to this device, 
you can bypass this step by using the pre-generated CSV files located at \texttt{path/to/MVE/data/csv\_files.tar.gz}.



\textbf{Software dependencies.}
Mobile measurements require Android NDK r23c (\href{https://github.com/android/ndk/wiki/Unsupported-Downloads}{\blue{download from Android GitHub repository}}) for cross-compilation, and Adreno OpenCL SDK v1.5 (available on \href{https://github.com/arkhadem/MVE/blob/main/tools/opencl-sdk-1.5.zip}{\blue{our GitHub repository}}) and Adreno OpenCL shared library (available on \href{https://github.com/arkhadem/MVE/tree/main/tools/libopencl}{\blue{our GitHub repository}}) for mobile GPU.
In addition to these, we need \href{https://developer.android.com/tools/adb}{\blue{Android Debug Bridge (\texttt{adb})}} to enable shell access to the phone.

\textbf{Installation.}
Install \texttt{adb}.
Download Android NDK r23c, Adreno OpenCL SDK v1.5, and Adreno OpenCL shared library, and extract them.
Run the following bash script to set up the cross-compilation infrastructure and compile the Arm Neon and Adreno GPU baselines.

\begin{figure*}[t]
    \centering
    \includegraphics[width=.90\textwidth]{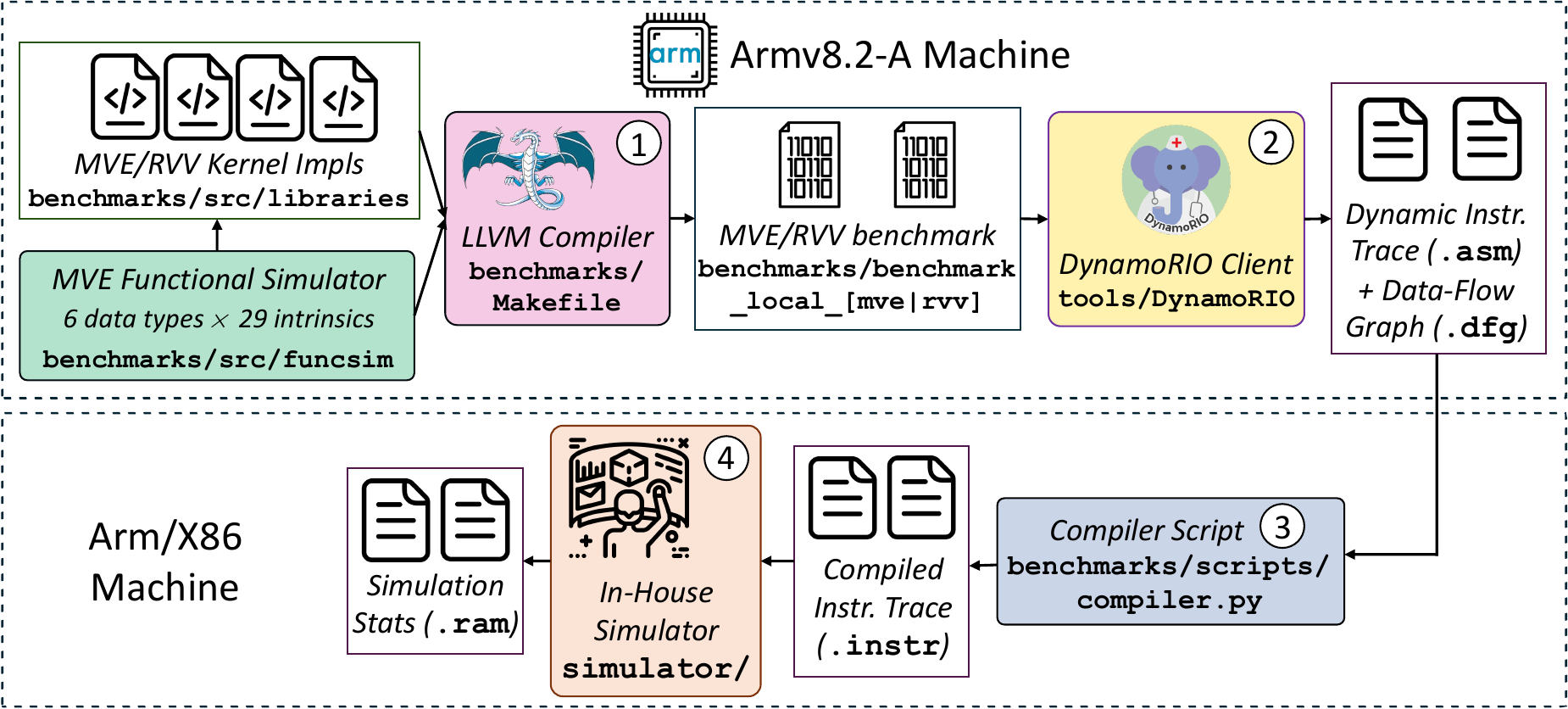}
    \caption{Simulation takes four steps. Steps 1 and 2 must be executed on an Armv8.2-A machine. Steps 3 and 4 can be performed on any machine.}
    \vspace{-4mm}
    \label{fig:sim}
\end{figure*}

\begin{lstlisting}
export ANDROID_NDK_ROOT=path/to/android-ndk-r23c
cp -r path/to/MVE/tools/opencl-sdk-1.5/inc/CL $ANDROID_NDK_ROOT/toolchains/llvm/prebuilt/linux-x86_64/sysroot/usr/include/
cp path/to/MVE/tools/libopencl/libOpenCL.so $ANDROID_NDK_ROOT/toolchains/llvm/prebuilt/linux-x86_64/lib/
cd path/to/MVE/benchmarks
make phone_neon -j
make phone_adreno -j
\end{lstlisting}

\textbf{Experiment Workflow.}
Connect the Samsung S10e device to your computer.
Enable the \href{https://developer.android.com/studio/debug/dev-options}{\blue{Developer Option and USB debugging}}.
Run the following provided scripts to automatically generate \texttt{neon.csv} and \texttt{adreno.csv} results.

\begin{lstlisting}
cd path/to/MVE/benchmarks
python scripts/profiler.py --core prime \
    --measurement performance --platform neon \
    --output neon.csv --verbose
python scripts/profiler.py  --core prime \
    --measurement performance --platform adreno \
    --output adreno.csv --verbose
\end{lstlisting}

\newpage

\subsection{Part 2 -- In-Cache Simulation}~\label{subsec:artifact_part2}

\textbf{Overview.}  
Figure~\ref{fig:sim} illustrates the four steps of the simulation process.  
Steps 1, 2, and 3 generate the instruction traces, while the final step 4 runs the simulations.  

\textbf{Hardware dependencies.}  
Steps 1 and 2 require a machine equipped with the Armv8.2-A ISA.  
The remaining steps 3 and 4 can be executed on any machine.  
If you do not have access to this certain machine, you may skip the first two steps.

\textbf{Software dependencies.}
We implemented a \href{https://github.com/arkhadem/MVE/blob/main/tools/DynamoRIO/samples/instrace.cpp}{\blue{DynamoRIO client}}~\cite{dynamorio} to instrument the benchmarks and dump the dynamic instruction traces.
We use \href{https://github.com/arkhadem/MVE/tree/main/simulator}{\blue{in-house simulator}} for in-cache computing simulations.

\textbf{Installation.}
Install the DynamoRIO client on a machine with Armv8.2-A ISA using the following script.
If you do not have access to such a machine, you can skip this step.

\begin{lstlisting}
cd path/to/MVE/tools/DynamoRIO/samples
mkdir build
cd build
cmake ..
make -j
\end{lstlisting}

To install our simulator, run the following script. Please note that building the simulator is mandatory, even if you plan to skip the first two steps of Part 2.

\begin{lstlisting}
cd path/to/MVE/simulator
bash make_all.sh
\end{lstlisting}

\textbf{Experiment Workflow.}
Simulation takes place in 4 steps using the following bash script.
\textit{Step 1} builds the \gpic{} and RVV baselines.
\textit{Step 2} runs the \gpic{} and RVV benchmarks and use the DynamoRIO client to generate a dynamic instruction trace (\texttt{.asm}) and a data-flow graph (\texttt{.dfg}) file for each kernel.
\textit{Step 3} compiles these files to the trace files for the simulator.
\textit{Step 4} runs the simulator and generates the simulation results (\texttt{.ram}).
To configure the number of concurrent simulations, use the \texttt{-j [NUM\_THREADS]} argument.
Finally, scripts are provided to parse the simulation results and generate the CSV files for any ISA (\gpic{} and RVV) and in-cache computing scheme (BS, BH, BP, AC) combinations.

\begin{lstlisting}
cd path/to/MVE/benchmarks
# Step 1: building the MVE and RVV kernels
make local_mve -j
make local_rvv -j
source ./set_env.sh
# Step 2: running DynamoRIO's instrumentation tool
python scripts/simulator.py --action benchmark \
        --directory ./data --verbose
# Step 3: compiling asm and dfg files into traces
python scripts/simulator.py --action compile \
        --directory ./data --verbose
# Step 4: running the simulations
python scripts/simulator.py --action simulate \
    --directory ./data -j [NUM_THREADS] --verbose
# Parsing the results
python scripts/simulator.py --action parse \
        --directory ./data --verbose
\end{lstlisting}

\textbf{Skipping Steps 1 and 2.}
Skip the first two steps by using the generated (\texttt{.asm}) and (\texttt{.dfg}) files.
First, ensure these files are downloaded (See Section~\ref{subsec:artifact_overview} -- How to access artifact?).
Next, perform the following operations to extract these files and resume \textit{Step 3}.

\begin{lstlisting}
cd path/to/MVE/benchmarks
tar -xzvf ../data/data_bs.tar.gz
tar -xzvf ../data/data_bh.tar.gz
tar -xzvf ../data/data_bp.tar.gz
tar -xzvf ../data/data_ac.tar.gz
# Resume Step 3
\end{lstlisting}

\subsection{Evaluation and expected results}

The expected results are in the pre-generated CSV files located at \texttt{path/to/MVE/data/csv\_files.tar.gz}. Use these files to fill out the first 10 sheets of the result Excel file, located at \texttt{path/to/MVE/data/Results.xlsx}. Completing these sheets will generate Figures~\ref{fig:performance_vs_cpus}(a),~\ref{figure:gpu_comparison},~\ref{figure:gemm_spmm},~\ref{fig:MVE_vs_RVV},~\ref{fig:MVE_vs_RVV_instr}, and~\ref{fig:MVE_vs_RVV_ALLIS}.


\newpage


\bibliographystyle{IEEEtranS}
\bibliography{refs}

\end{document}